\documentclass[longauth]{aa}

\usepackage{txfonts}
\usepackage{graphicx}
\usepackage{natbib}
\bibpunct{(}{)}{;}{a}{}{,}

\newcommand{\nwd}{5\,926}
\newcommand{\ewca}{{\rm EW}_{J0395}}
\newcommand{\pca}{p_{\rm Ca}}

\begin{document}

\title{J-PLUS: The fraction of calcium white dwarfs along the cooling sequence
\thanks{The catalog with the parameters of the analyzed white dwarfs is available both on the \texttt{jplus.WhiteDwarf} table at J-PLUS database and 
at the CDS via anonymous ftp to \url{cdsarc.u-strasbg.fr} (130.79.128.5)
or via \url{http://cdsarc.u-strasbg.fr/viz-bin/cat/J/A+A/XXX/AXXX}.}
}

\author{C.~L\'opez-Sanjuan\inst{\ref{CEFCA}}
\and P.-E.~Tremblay\inst{\ref{warwick}}
\and M.~W.~O'Brien\inst{\ref{warwick}}
\and D.~Spinoso\inst{\ref{tsinghua},\ref{CEFCA2}}
\and A.~Ederoclite\inst{\ref{CEFCA}}
\and H.~V\'azquez Rami\'o\inst{\ref{CEFCA}}
\and A.~J.~Cenarro\inst{\ref{CEFCA}}
\and A.~Mar\'{\i}n-Franch\inst{\ref{CEFCA}}
\and T.~Civera\inst{\ref{CEFCA2}}
\and J.~M.~Carrasco\inst{\ref{ICCUB},\ref{UB},\ref{IEEC}}
\and B.~T.~G\"ansicke\inst{\ref{warwick}}
\and N.~P.~Gentile~Fusillo\inst{\ref{trieste}}
\and A.~Hern\'an-Caballero\inst{\ref{CEFCA}}
\and M.~A.~Hollands\inst{\ref{sheffield}}
\and A.~del Pino\inst{\ref{CEFCA}}
\and H.~Dom\'{\i}nguez S\'anchez\inst{\ref{CEFCA}}
\and J.~A.~Fern\'andez-Ontiveros\inst{\ref{CEFCA}}
\and F.~M.~Jim\'enez-Esteban\inst{\ref{CAB}}
\and A.~Rebassa-Mansergas\inst{\ref{UPC},\ref{IEEC}}
\and L.~Schmidtobreick\inst{\ref{ESO}}
\and R.~E.~Angulo\inst{\ref{DIPC},\ref{ikerbasque}}
\and D.~Crist\'obal-Hornillos\inst{\ref{CEFCA2}}
\and R.~A.~Dupke\inst{\ref{ON},\ref{MU},\ref{Alabama}}
\and C.~Hern\'andez-Monteagudo\inst{\ref{IAC},\ref{ULL}}
\and M.~Moles\inst{\ref{CEFCA2}}
\and L.~Sodr\'e Jr.\inst{\ref{USP}}
\and J.~Varela\inst{\ref{CEFCA2}}
}

\institute{Centro de Estudios de F\'{\i}sica del Cosmos de Arag\'on (CEFCA), Unidad Asociada al CSIC, Plaza San Juan 1, 44001 Teruel, Spain\\\email{clsj@cefca.es}\label{CEFCA}   
        \and
        Department of Physics, University of Warwick, Coventry, CV4 7AL, UK\label{warwick} 
        \and
        Department of Astronomy, 6th floor, MongManWai Building, Tsinghua University, Beijing 100084, People's Republic of China\label{tsinghua}
        \and
        Centro de Estudios de F\'{\i}sica del Cosmos de Arag\'on (CEFCA), Plaza San Juan 1, 44001 Teruel, Spain\label{CEFCA2}
        \and
        Institut de Ci\`encies del Cosmos (ICCUB), Universitat de Barcelona (UB), Mart\'{\i} i Franqu\`es 1, 08028 Barcelona, Spain\label{ICCUB}
        \and
        Departament de F\'{\i}sica Qu\`antica i Astrof\'{\i}sica (FQA), Universitat de Barcelona (UB), Mart\'{\i} i Franqu\`es 1, 08028 Barcelona, Spain\label{UB}
        \and
        Institut d'Estudis Espacials de Catalunya (IEEC), Esteve Terradas, 1, Edifici RDIT, Campus PMT-UPC, 08860 Castelldefels (Barcelona), Spain\label{IEEC}
        \and
        Department of Physics, Universita’ degli Studi di Trieste, Via A. Valerio 2, 34127, Trieste, Italy\label{trieste}
        \and
        Department of Physics and Astronomy, University of Sheffield, Sheffield, S3 7RH, UK\label{sheffield}
        \and
        Centro de Astrobiolog\'{\i}a (CAB), CSIC-INTA, Camino Bajo del Castillo s/n, 28692, Villanueva de la Ca\~nada, Madrid, Spain\label{CAB}
        \and
        Departament de F\'{\i}sica, Universitat Polit\`ecnica de Catalunya, c/Esteve Terrades 5, 08860 Castelldefels, Spain\label{UPC}
        \and
        European Southern Observatory, Casilla 19001, Santiago 19, Chile\label{ESO}
        \and
        Donostia International Physics Centre (DIPC), Paseo Manuel de Lardizabal 4, 20018 Donostia-San Sebastián, Spain\label{DIPC}
        \and
        IKERBASQUE, Basque Foundation for Science, 48013, Bilbao, Spain\label{ikerbasque}
        \and
        Observat\'orio Nacional - MCTI (ON), Rua Gal. Jos\'e Cristino 77, S\~ao Crist\'ov\~ao, 20921-400 Rio de Janeiro, Brazil\label{ON}
        \and
        University of Michigan, Department of Astronomy, 1085 South University Ave., Ann Arbor, MI 48109, USA\label{MU}
        \and
        University of Alabama, Department of Physics and Astronomy, Gallalee Hall, Tuscaloosa, AL 35401, USA\label{Alabama}
        \and
        Instituto de Astrof\'{\i}sica de Canarias, La Laguna, 38205, Tenerife, Spain\label{IAC}
        \and
        Departamento de Astrof\'{\i}sica, Universidad de La Laguna, 38206, Tenerife, Spain\label{ULL}
        \and
        Instituto de Astronomia, Geof\'{\i}sica e Ci\^encias Atmosf\'ericas, Universidade de S\~ao Paulo, 05508-090 S\~ao Paulo, Brazil\label{USP}
}

\date{Received 23 June 2024 / Accepted X X X}

\abstract
{}
{We used the Javalambre Photometric Local Universe Survey (J-PLUS) second data release (DR2) photometry in twelve optical bands over $2\,176$ deg$^2$ to estimate the fraction of white dwarfs with presence of \ion{Ca}{ii} H+K absorption along the cooling sequence.}
{We compared the J-PLUS photometry against metal-free theoretical models to estimate the equivalent width in the $J0395$ passband of $10$ nm centered at $395$ nm ($\ewca$), a proxy to detect calcium absorption. A total of $4\,399$ white dwarfs with effective temperature within $30\,000 > T_{\rm eff} > 5\,500$ K and mass $M > 0.45$ $M_{\odot}$ were analyzed. Their $\ewca$ distribution was modeled using two populations, corresponding to polluted and non-polluted systems, to estimate the fraction of calcium white dwarfs ($f_{\rm Ca}$) as a function of $T_{\rm eff}$. The probability for each individual white dwarf of presenting calcium absorption, $\pca$, was also computed.}
{The comparison of $\ewca$ with both the measured Ca/He abundance and the identification of metal pollution from spectroscopy shows that $\ewca$ correlates with the presence of \ion{Ca}{ii} H+K absorption. The fraction of calcium white dwarfs changes along the cooling sequence, increasing from $f_{\rm Ca} \approx 0$ at $T_{\rm eff} = 13\,500$ K to $f_{\rm Ca} \approx 0.15$ at $T_{\rm eff} = 5\,500$ K. This trend reflects the selection function of calcium white dwarfs in the optical. We compare our results with the fractions derived from the 40 pc spectroscopic sample and from Sloan Digital Sky Survey (SDSS) spectra. The trend found in J-PLUS observations is also present in the 40 pc sample, however SDSS shows a deficit of metal-polluted objects at $T_{\rm eff} < 12\,000$ K. Finally, we found $39$ white dwarfs with $\pca > 0.99$. Twenty of them have spectra presented in previous studies, whereas we obtained follow-up spectroscopic observations for six additional targets. These $26$ objects were all confirmed as metal-polluted systems.}
{The J-PLUS optical data provide a robust statistical measurement for the presence of \ion{Ca}{ii} H+K absorption in white dwarfs. We find a $15 \pm 3$\% increase in the fraction of calcium white dwarfs from $T_{\rm eff} = 13\,500$~K to $5\,500$~K, which reflects their selection function in the optical from the total population of metal-polluted systems.}


\titlerunning{J-PLUS. The fraction of calcium white dwarfs along the cooling sequence}

\authorrunning{L\'opez-Sanjuan et al.}

\maketitle

\section{Introduction}\label{sec:intro}
\begin{figure*}[t]
\centering
\resizebox{0.49\hsize}{!}{\includegraphics{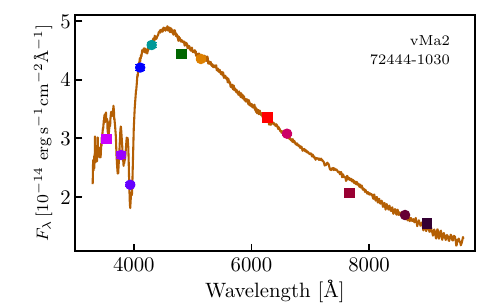}}
\resizebox{0.49\hsize}{!}{\includegraphics{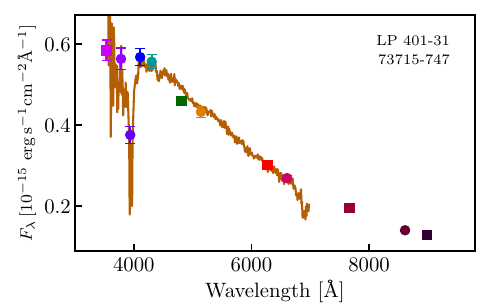}}
\caption{Spectral energy distribution of van Maanen 2 ({\it left panel}) and LP $401$-$31$ ({\it right panel}), two metal-polluted white dwarfs. The colored symbols in both panels are the $3$ arcsec diameter aperture photometry corrected to total flux from J-PLUS (squares for broad bands, $ugriz$; circles for medium bands, $J0378$, $J0395$, $J0410$, $J0430$, $J0515$, $J0660$, and $J0861$). The passband $J0395$ is the most sensitive to the presence of \ion{Ca}{ii} H+K absorption. The brown line shows the spectrum of the source from the {\it Gaia} spectro-photometric standard stars survey \citep[][{\it left panel}]{gaia_spss_i,gaia_spss_v} and \citet[][{\it right panel}]{limoges15}. The unique J-PLUS identification, composed by the \texttt{TILE\_ID} of the reference $r-$band image and the \texttt{NUMBER} assigned by \texttt{SExtractor} to the source, is reported in the panels for reference.}
\label{fig:intro}
\end{figure*}

The presence of metals in the atmosphere of cool white dwarfs is ascribed to the accretion of material from the ancient planetary system (disrupted planets, asteroids, comets) of the degenerate star \citep[e.g.][]{jura03,jura08,farihi10}. The presence of metals is indicated by the detection of absorption lines in the white dwarf spectrum. Modeling the metal lines with a careful treatment of diffusion timescales from theory permits to derive the chemical abundances of the planetary debris \citep[e.g.][]{jura14,hollands17,turner20,blouin20} and the accretion rates \citep[e.g.][]{koester06,hollands18dz,blouin22}. Thus, the study of metal-polluted white dwarfs provides clues about the formation and evolution of exoplanetary systems.

The fraction of metal-polluted white dwarfs can be as high as $30-50$\% when the white dwarf population is analyzed with ultraviolet spectroscopy \citep{koester14} or high-resolution ($R \gtrsim 20\,000$) optical spectroscopy \citep{zuckerman03,zuckerman10,koester05}. However, the discovery of these systems has been mainly serendipitous from Sloan Digital Sky Survey (SDSS) medium-resolution ($R \sim 2\,000$) optical spectroscopy \citep{kleinman13,kepler15,kepler16,kepler19}, which is affected by non-trivial selection effects. This hampers the interpretation and representativeness of the measured chemical abundances and accretion rates from the roughly one thousand metal-polluted white dwarfs available in the literature \citep{coutu19}.

The main selection bias for metal-polluted systems is the ability to detect absorption lines. The most prominent feature in the optical corresponds to the \ion{Ca}{ii} H+K line, whose equivalent width (EW) depends on both the effective temperature ($T_{\rm eff}$) and the atmospheric calcium abundance. This is convolved with the resolution and the signal-to-noise ($S/N$) of the spectrum, defining a minimum EW for detection which effectively sets an observational limit for Ca in white dwarfs. In addition, the target selection for the spectroscopic observations also impacts the calcium detectability. In this context, white dwarfs with the presence of calcium absorption in the optical, also known as calcium white dwarfs, are a sub-sample of the complete metal-polluted white dwarf population.. 

A novel way to analyze the incidence of calcium white dwarfs is using photometric surveys with passbands sensitive to the presence of calcium absorption, such as the Javalambre Photometric Local Universe Survey (J-PLUS, $12$ optical filters including the $J0395$ passband of $100~\AA$ width centered at $3\,939~\AA$, Fig.~\ref{fig:intro}; \citealt{cenarro19}), the Southern Photometric Local Universe Survey (S-PLUS, with the same filter system as J-PLUS; \citealt{splus}), Pristine (a unique CaHK filter of $98~\AA$ width centered at $3\,952~\AA$; \citealt{pristine}), and the Javalambre Physics of the Accelerating Universe Astrophysical Survey ($54$ contiguous filters of $140~\AA$ in the optical, including two passbands centered at $3\,900$ and $4\,000~\AA$; \citealt{jpas,minijpas}). The medium band photometry in these surveys has typically a poorer sensitivity than spectroscopy for the detection of \ion{Ca}{ii} H+K absorption, but it enables an homogeneous selection over the surveyed area.

In this work, we provide observational constraints to the calcium white dwarf fraction detected in the optical at $30\,000 > T_{\rm eff} > 5\,500$~K as a proxy to understand the selection of metal-polluted systems. We used the J-PLUS second data release (DR2; \citealt{clsj21zsl}) covering $2\,176$ deg$^2$ to supplement the white dwarf catalog presented in \citet[GF21 hereafter]{GF21}, which is based on {\it Gaia} EDR3 \citep{gaia,gaiaedr3}. The J-PLUS photometric results are complemented with the spectroscopic data of the $40$ pc sample \citep{40pciii} and the SDSS DR16 spectra \citep{sdss_dr16} compiled in the GF21 catalog.

This paper is organized as follows. In Sect.~\ref{sec:data} we detail the J-PLUS photometric data and the reference white dwarf catalog used in our analysis. The estimation of the EW in the $J0395$ filter and the atmospheric parameters of the white dwarfs are presented in Sect.~\ref{sec:methods}. We compare the estimated EWs and effective temperatures derived from photometry with spectroscopic measurements from the literature in Sect.~\ref{sec:ewca_test}. The evolution in the fraction of calcium white dwarfs with the effective temperature by a Bayesian analysis is presented in Sect.~\ref{sec:fca}. The comparison with the spectroscopic samples and the discussion about the selection function in the detection of metal-polluted white dwarfs are detailed in Sect.\ref{sec:selec_spec}. We devote Sect.~\ref{sec:pca} to derive the individual probability of having calcium absorption and present the spectroscopic follow-up of our most probable candidates. Finally, the summary and conclusions are presented in Sect.~\ref{sec:conclusions}. All magnitudes are expressed in the AB system \citep{oke83}.

\section{Data}\label{sec:data}
\subsection{J-PLUS photometric data}\label{sec:jplus}
J-PLUS\footnote{\url{www.j-plus.es}} is being conducted at the Observatorio Astrof\'{\i}sico de Javalambre (OAJ, Teruel, Spain; \citealt{oaj}) using the 83\,cm Javalambre Auxiliary Survey Telescope (JAST80) and T80Cam, a panoramic camera of 9.2k $\times$ 9.2k pixels. The T80cam camera at JAST80 provides a $2\deg^2$ field of view (FoV) with a pixel scale of 0.55 arsec pix$^{-1}$ \citep{t80cam}. The J-PLUS photometric system (Table~\ref{tab:JPLUS_filters}) comprises five SDSS-like ($ugriz$) and seven medium band filters centered at key stellar features, such as the $4\,000~\AA$ break ($J0378$, $J0395$, $J0410$, and $J0430$), the Mg $b$ triplet ($J0515$), H$\alpha$ at rest-frame ($J0660$), or the calcium triplet ($J0861$). The most relevant passband in this work is the $J0395$, the third filter in ascending wavelength order, that is sensitive to the presence of \ion{Ca}{ii} H+K absorption as illustrated in Fig.~\ref{fig:intro}. The observational strategy, image reduction, and scientific goals of the survey are summarized in \citet{cenarro19}.

The J-PLUS DR2 comprises $1\,088$ pointings ($2\,176$ deg$^2$) observed, reduced, and calibrated in all survey bands \citep{clsj21zsl}. The limiting magnitudes (5$\sigma$, 3 arsec aperture) are $\sim 22$ mag in $g$ and $r$ passbands, and $\sim 21$ mag in the other passbands. The median point spread function (PSF) full width at half maximum (FWHM) in the $r$ band is $1.1$ arcsec. Source detection was done in the $r$-band images using \texttt{SExtractor} \citep{sextractor}, and the fluxes measured in the twelve J-PLUS images using the position of the detected sources. Those objects close to bright stars, near to the borders of the images, or affected by artifacts were masked from the initial footprint, providing a high-quality area of $1\,941$ deg$^2$. The DR2 is publicly available on the J-PLUS website\footnote{\url{www.j-plus.es/datareleases/data_release_dr2}}.

We used 3 arcsec diameter aperture photometry to analyze the white dwarf population. The observed fluxes were stored in the vector $\vec{f} = \{ f_j \}$, and their uncertainties in the vector $\sigma_{\vec{f}} = \{\sigma_j\}$, where the index $j$ runs over the J-PLUS passbands (Table~\ref{tab:JPLUS_filters}). The uncertainty vector includes the errors from photon counting, sky background, and photometric calibration \citep{clsj21zsl}. We note that J-PLUS covers the Milky Way halo, so crowding is not affecting the quality of the $3$ arcsec photometry.

\begin{table} 
\caption{J-PLUS passbands, including filter transmission, CCD efficiency, telescope optics, and atmosphere.}
\label{tab:JPLUS_filters}
\centering 
        \begin{tabular}{l c c c}
        \hline\hline\rule{0pt}{3ex} 
        Passband   & Index & Effective &   FWHM\\
                   &  $j$  & wavelength  &  \\
                   &       & [nm]             &   [nm]\\
        \hline\rule{0pt}{2ex}
        \!$u$     & 1   &353.3  &  50.8     \\ 
        $J0378$   & 2   &378.2  &  16.8     \\ 
        $J0395$   & 3   &393.9  &  10.0     \\ 
        $J0410$   & 4   &410.8  &  20.0     \\ 
        $J0430$   & 5   &430.3  &  20.0     \\ 
        $g$       & 6   &479.0  & 140.9     \\ 
        $J0515$   & 7   &514.1  &  20.0     \\ 
        $r$       & 8   &625.7  & 138.8     \\ 
        $J0660$   & 9   &660.4  &  13.8     \\ 
        $i$       & 10  &765.6  & 153.5     \\ 
        $J0861$   & 11  &861.0  &  40.0     \\ 
        $z$       & 12  &896.5  & 140.9     \\ 
        \hline 
\end{tabular}
\end{table}

\subsection{White dwarf catalog}\label{sec:wdcat}
We based our study on the J-PLUS + {\it Gaia} white dwarf catalog defined by \citet{clsj22pda}. This sample was extracted from the {\it Gaia} DR3 catalog of white dwarfs presented in GF21 and complemented with the J-PLUS DR2 twelve-band photometry. The sample comprises $5\,926$ white dwarfs with probability $P_{\rm WD} > 0.75$ in GF21, $r \leq 19.5$ mag, and parallax $1 < \varpi < 100$ mas. We refer the reader to \citet{clsj22pda} for an extended description of the sample, including the Bayesian classification of the sources in H- and He-dominated atmospheres.

\begin{figure*}[t]
\centering
\resizebox{0.49\hsize}{!}{\includegraphics{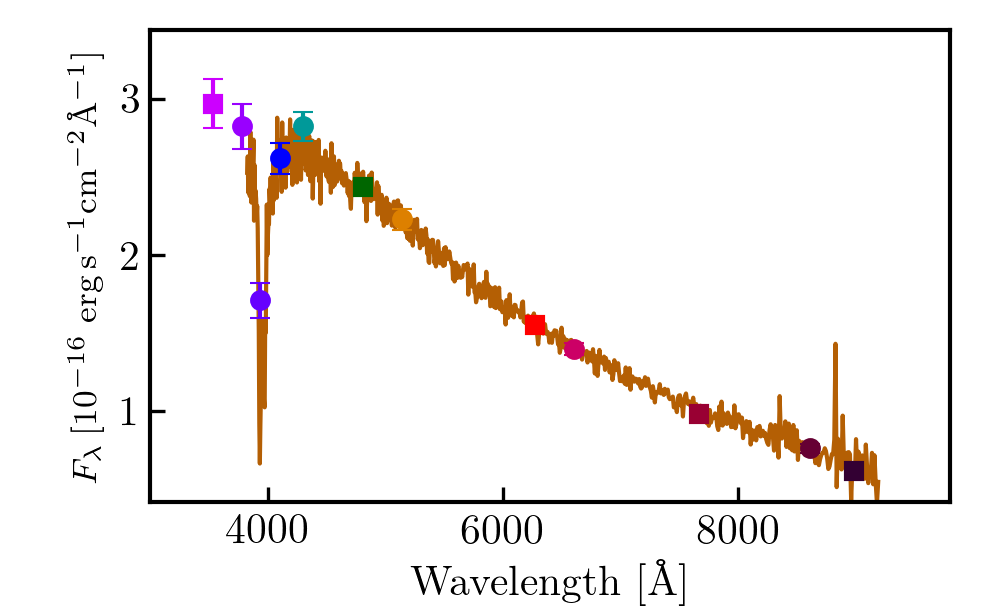}}
\resizebox{0.49\hsize}{!}{\includegraphics{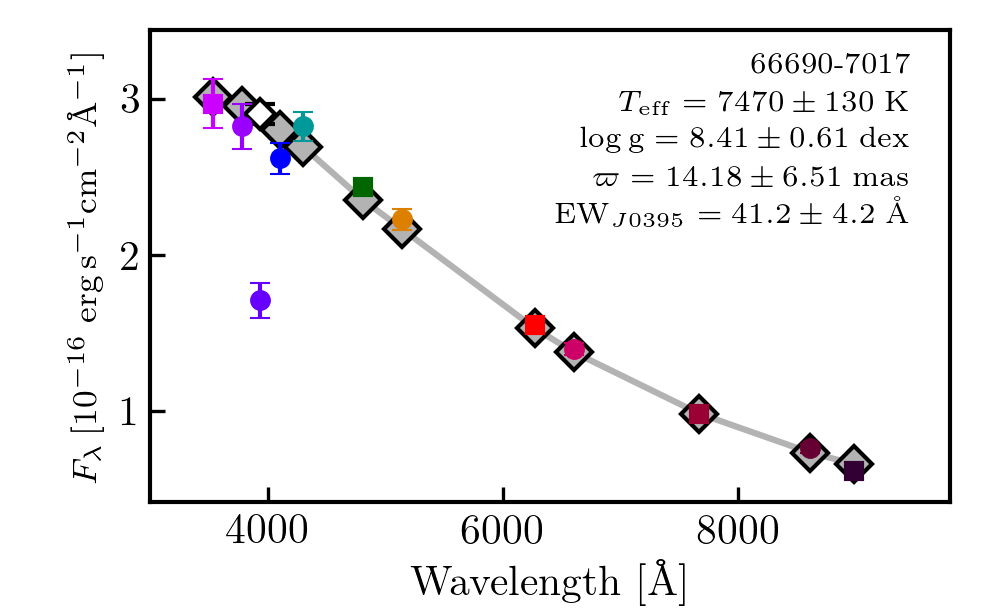}}
\caption{Illustrative example for the estimation of the $J0395$ equivalent width, ${\rm EW}_{J0395}$, in WD $1529$+$428$, a DZ with spectroscopic information from SDSS ({\it left panel}). Symbols are showing the J-PLUS photometry as in Fig.~\ref{fig:intro}. The gray diamonds connected with a solid line in the {\it right panel} depict the best-fit solution to J-PLUS photometry but the $J0395$ passband, with the derived parameters and their uncertainties labeled: effective temperature ($T_{\rm eff}$), surface gravity ($\log {\rm g}$) and parallax ($\varpi$). The white diamond represents the expected continuum flux in the $J0395$ filter in absence of polluting metals from the fitting process, see Eq.~(\ref{eq:pdfvar}). The estimated $J0395$ equivalent width is also presented in the panel.}
\label{fig:example}
\end{figure*}

\subsection{Spectroscopic information}\label{sec:spec}
The GF21 catalog also gathers the available spectroscopic information from SDSS DR16 \citep{sdss_dr16}, including an updated spectral classification of the white dwarfs. The main types in the catalog are DA (hydrogen lines), DB (helium lines), DC (featureless continuum), DZ (metal lines), and DQ (carbon lines). We found $1\,393$ sources with $S/N > 5$ in common with our white dwarf sample, including $59$ objects with signs of metal pollution.

We also searched for objects in \citet{coutu19}. The authors perform a detailed abundance analysis of $1\,023$ metal-polluted DZ and DBZ white dwarfs, providing an estimation of their effective temperature and the calcium-over-helium abundance, noted Ca/He. There are $44$ common sources with $15\,000 > T_{\rm eff} > 5\,000$ K and $-11 < \log {\rm Ca/He} < -7$ dex.

Finally, we also used the so-called 40 pc sample presented in \citet{40pciv}. This volume-limited sample was selected from GF21 candidates and comprises $1\,076$ white dwarfs with both dedicated spectroscopy and ancillary data \citep{limoges15,40pci,40pcii,40pciii}. There are $119$ metal-polluted white dwarfs in the sample. We refer the reader to the aforementioned papers for an extensive discussion of the 40 pc sample.

\section{Estimation of the $J0395$ equivalent width}\label{sec:methods}
We aim to detect the presence of calcium absorption in the sample of $\nwd$ white dwarfs presented in \citet{clsj22pda}. A possible approach is to analyze the J-PLUS photometry using a set of theoretical models with and without metal-polluted atmospheres. This provides the likelihood for the presence of metals, at the cost of increasing the number of free parameters in the fitting procedure with the abundances of the main polluting elements (i.e. Ca, Mg, and Fe). The relative abundances of these elements can be fixed to those measured in Solar System chondrites \citep[e.g.][]{lodders03}, reducing the extra parameters in the analysis to the calcium abundance.

Even in the simplified case, there are two drawbacks: the relative low fraction of metal-polluted systems with strong optical features, ranging from $5$\% to $15$\% \citep{hollands18,40pcii,40pciii}, and the larger sensitivity of the $J0395$ filter to the presence of calcium absorption with respect the other J-PLUS passbands. In this scenario, half of the white dwarfs would have a larger likelihood of being metal-polluted just because of the random errors in the $J0395$ photometry of normal, non-polluted systems. In other words, we face a highly degenerate problem.

To solve this statistical challenge, a proper prior is needed in the fraction of calcium white dwarfs as a function of the white dwarf properties, e.g. the effective temperature. This is precisely the main goal of the present paper. We circumvented this circular argument by simplifying the analysis as much as possible, defining a proxy for the presence of calcium absorption avoiding theoretical modeling of metal-polluted white dwarfs. The adopted proxy was the equivalent width in the $J0395$ filter:
\begin{equation}
    {\rm EW}_{J0395} = 100 \times \bigg( 1 - \frac{f_{J0395}}{f_{J0395}^{\rm cont}} \bigg)\ \ [\AA],\label{eq:ewj0395}
\end{equation}
where $f_{J0395}$ is the observed flux in the $J0395$ passband and $f_{J0395}^{\rm cont}$ is the expected continuum flux in absence of metal pollution. The estimation of the continuum flux is detailed in the next section and the goodness of $\ewca$ as a proxy for the presence of calcium is shown in Sect.~\ref{sec:ewca_test}.

\subsection{Estimation of the continuum flux in $J0395$}
The first step to determine the expected continuum flux in absence of polluting metals for each white dwarf in the sample is the estimation of the following probability density function (PDF),
\begin{equation}
    {\rm PDF}\,(t,\theta\,|\,\vec{f}, \vec{\sigma}_{\vec{f}}) \propto \mathcal{L}\,(\,\vec{f}\,|\,t,\theta,\sigma_{\vec{f}}) \times P\,(\theta),
\end{equation}
where $\theta$ are the parameters in the fitting, $t$ are the different non-polluted atmospheric compositions considered in our analysis, $\mathcal{L}$ is the likelihood of the data for a given set of parameters and composition, and $P$ is the prior probability. The parameters in the fitting were $\theta = \{T_{\rm eff}, \log {\rm g}, \varpi \}$, corresponding to effective temperature, surface gravity, and parallax. We explored two atmospheric compositions, corresponding to hydrogen- and helium-dominated atmospheres, $t = \{ {\rm H}, {\rm He} \}$. The final PDF is normalized to one by definition.

We defined the likelihood of the data given a set of parameters and composition as
\begin{equation}
    \mathcal{L}\,(\,\vec{f}\,|\,t,\theta,{\boldmath \sigma}_{\vec{f}}) = \prod_{\substack{j = 1\\j \neq 3}}^{12} P_{\rm G}\,(f_j\,|\,f_{j}^{\rm mod}, \sigma_{j}),
\end{equation}
where the index $j$ runs over the J-PLUS passbands (Table~\ref{tab:JPLUS_filters}) excluding $J0395$, the function $P_{\rm G}$ defines a Gaussian probability distribution, 
\begin{equation}
P_{\rm G}\,(x\,|\,\mu,\sigma) = \frac{1}{\sqrt{2\pi}\,\sigma}\ {\rm exp}\Big[-\frac{(x - \mu)^2}{2\sigma^2}\Big],
\end{equation}
and the model flux was estimated as
\begin{equation}
    f_{j}^{\rm mod}\,(t,\theta)= \bigg( \frac{\varpi}{100} \bigg)^2\,F_{t,j}\,(T_{\rm eff},\log {\rm g})\,10^{-0.4\,k_j\,E(B-V)}\,10^{0.4\,C^{\rm aper}_j},
\end{equation}
where $F_{t,j}$ is the theoretical absolute flux emitted by a white dwarf of type $t$ located at 10 pc distance, $k_j$ is the extinction coefficient of the passband, $E(B-V)$ is the color excess of the white dwarf, and $C^{\rm aper}_j$ is the aperture correction needed to translate the observed $3$ arcsec diameter magnitudes to total magnitudes. The estimation of the color excess and the aperture correction are detailed in \citet{clsj22pda}.

The H-dominated atmospheres were described with pure-H models ($t = {\rm H}$, \citealt{tremblay11,tremblay13}), while mixed models with H/He = $10^{-5}$ at $T_{\rm eff} > 6\,500$ K and pure-He models at $T_{\rm eff} < 6\,500$ K were used to describe He-dominated atmospheres ($t = {\rm He}$, \citealt{cukanovaite18, cukanovaite19}). We assumed the mass-radius relation of \citet{fontaine01} for thin (He-atmospheres) and thick (H-atmospheres) hydrogen layers. A discussion about these models is presented elsewhere \citep[e.g.][]{bergeron19,GF20,40pcii}.

A prior probability for the parallax was used,
\begin{equation}
    P\,(\varpi) = P_{\rm G}\,(\varpi\,|\,\varpi_{\rm DR3}, \sigma_{\varpi}),
\end{equation}
where $\varpi_{\rm  DR3}$ and $\sigma_{\varpi}$ are the parallax and its error obtained from ${\it Gaia}$ DR3 \citep{gaiaedr3,lindegren21a}. The published values of the parallax were corrected by the {\it Gaia} zero-point offset following the prescription by \citet{lindegren21b}. Finally, a flat prior for $T_{\rm eff}$ and $\log {\rm g}$ was assumed.

At this stage, we had the PDF of the atmospheric parameters and the composition for each white dwarf. The photometry in the $J0395$ passband was not used in the fitting process, and the PDF of the expected continuum flux was estimated as 
\begin{equation}
    {\rm PDF}\,(f_{J0395}) = \sum_t \int {\rm PDF}\,(t,\theta) \times \delta[f_{J0395} - f_{J0395}^{\rm mod}\,(t,\theta)]\,{\rm d}\theta,\label{eq:pdfvar}
\end{equation}
where $\delta$ is the Dirac delta function. The final value of $f_{J0395}^{\rm cont}$ and its uncertainty were estimated as the median and the dispersion of the best-fit Gaussian to the distribution ${\rm PDF}(f_{J0395})$. We found that this process provides a proper description of the posterior and permits the measurement of the $J0395$ equivalent width using Eq.~(\ref{eq:ewj0395}).

The measurement process is illustrated in Fig.~\ref{fig:example} with the J-PLUS source $66690$-$7017$ (WD $1529$+$428$). This white dwarf is classified as DZ \citep{eisenstein06_dr4} and presents an intense \ion{Ca}{ii} H+K absorption, as highlighted by the SDSS spectrum. The measured equivalent width was $\ewca = 41.2 \pm 4.2$ \AA, or a 10$\sigma$ detection for the presence of polluting metals using J-PLUS photometry.

\begin{figure*}[t]
\centering
\resizebox{0.49\hsize}{!}{\includegraphics{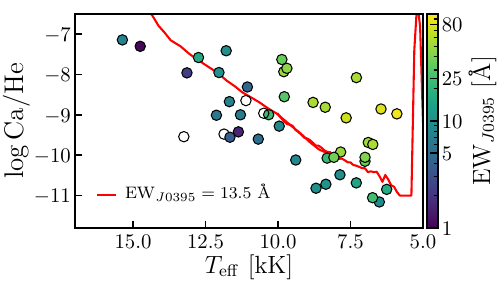}}
\resizebox{0.49\hsize}{!}{\includegraphics{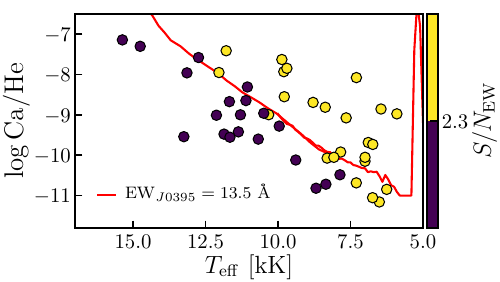}}
\caption{Calcium relative abundance, $\log {\rm Ca/He}$, as a function of effective temperature, $T_{\rm eff}$, from \citet{coutu19} for the $44$ white dwarfs in common with the J-PLUS + {\it Gaia} sample (colored circles). The solid line in both panels shows the expected detection limit for ${\rm EW}_{J0395} = 13.5~\AA$ derived from theoretical models with $\log {\rm g} = 8$. {\it Left panel}: The color scale represents ${\rm EW}_{J0395}$, with white symbols having ${\rm EW}_{J0395} < 0~\AA$. {\it Right panel}: The color scale represents signal-to-noise in the $J0395$ equivalent width with two discrete possibilities, being larger or lower than $S/N_{\rm EW} = 2.3$. This value corresponds to a $99$\% detection or ${\rm EW}_{J0395} \approx 13.5~\AA$ given the uncertainties in the measurements.}
\label{fig:cahe_teff}
\end{figure*}

\begin{figure*}[t]
\centering
\resizebox{0.49\hsize}{!}{\includegraphics{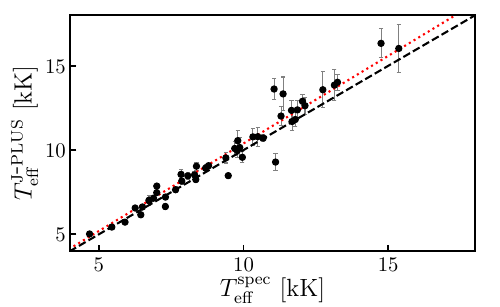}}
\resizebox{0.49\hsize}{!}{\includegraphics{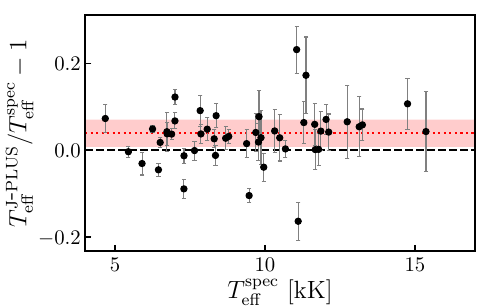}}
\caption{Effective temperature from J-PLUS photometry, $T_{\rm eff}^{\rm J\textrm{-}PLUS}$, as a function of the spectroscopic value derived by \citet{coutu19}, $T_{\rm eff}^{\rm spec}$ ({\it left panel}). The black dashed line marks the one-to-one relation and the red dotted line a 4\% overestimation from J-PLUS with respect to the spectroscopic temperature. {\it Right panel}: Relative difference between J-PLUS and spectroscopic $T_{\rm eff}$ as a function of $T_{\rm eff}^{\rm spec}$. The black dashed line marks a zero difference and the red dotted line a $4$\% overestimation. The red area encloses 68\% of the points and it is equivalent to a $3$\% dispersion.}
\label{fig:teffspec}
\end{figure*}

\subsection{White dwarf parameters and summary statistics}
In addition to $\ewca$, we also estimated the effective temperature and the mass of the analyzed white dwarfs. We marginalized over surface gravity, parallax, and composition to obtain ${\rm PDF}\,(T_{\rm eff})$, and used the mass-weighted version of Eq.~(\ref{eq:pdfvar}) to estimate the mass posterior,
\begin{equation}
    {\rm PDF}\,(M) = \sum_t \int {\rm PDF}\,(t,\theta) \times \delta[M - M^{\rm mod}\,(t,\theta)]\,{\rm d}\theta.\label{eq:pdfmass}
\end{equation}
The retrieved parameters and their uncertainties were again the median and the dispersion of the best-fit Gaussian to the posterior distributions.

The catalog with the parameters used in the present work is publicly available at the J-PLUS database and the Centre de Donn\'ees astronomiques de Strasbourg\footnote{\url{https://cds.u-strasbg.fr}} (CDS). The description of the catalog is presented in Appendix~\ref{app:data}.

\begin{figure*}[t]
\centering
\resizebox{0.49\hsize}{!}{\includegraphics{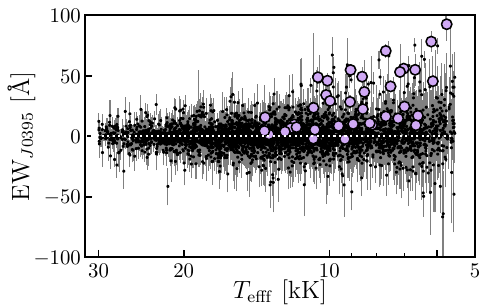}}
\resizebox{0.49\hsize}{!}{\includegraphics{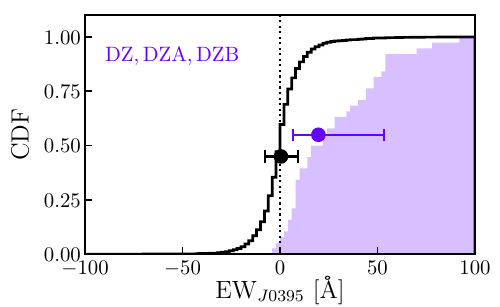}}\\
\resizebox{0.49\hsize}{!}{\includegraphics{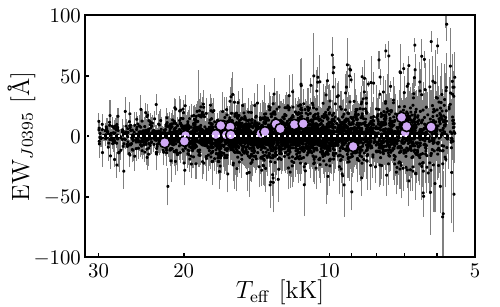}}
\resizebox{0.49\hsize}{!}{\includegraphics{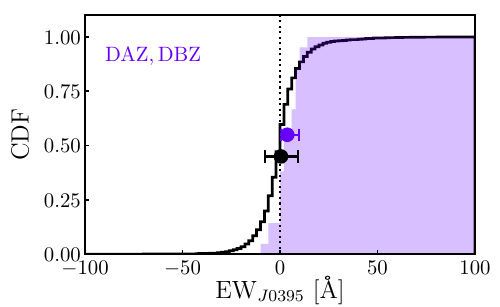}}
\caption{{\it Left panels} : $J0395$ equivalent width, $\ewca$, as a function of the effective temperature from J-PLUS photometry (black bullets). The top panels show the $38$ white dwarfs with a metal dominant type (DZ, DZA, DZB, etc.) and the bottom panels the $21$ white dwarfs with a non-dominant metal type (DAZ, DBZ, etc.) as purple circles. The spectroscopic type is from GF21 for SDSS DR16 spectra with $S/N \geq 5$. {\it Right panels} : Cumulative distribution function (CDF) in ${\rm EW}_{J0395}$ for the total sample (black) and the metal-polluted sample (purple). The black and purple circles with error bars depict the median and the [16,84] percentiles of the distribution.}
\label{fig:ewdist}
\end{figure*}

\section{Correlation of $\ewca$ with calcium abundance}\label{sec:ewca_test}
In this section, the reliability of $\ewca$ as a proxy for the presence of calcium absorption in white dwarfs is evaluated. We compare the calcium abundance and the spectroscopic temperature from \citet{coutu19} with their photometric counterparts in Sects.~\ref{sec:cahe_teff_spec} and \ref{sec:teff_spec}. The $\ewca$ of those white dwarfs with metal pollution from the GF21 classification using SDSS DR16 data are presented in Sect.~\ref{sec:caew_dist}.

\subsection{Ca/He abundance versus effective temperature}\label{sec:cahe_teff_spec}
We found $44$ sources in common between our J-PLUS + {\it Gaia} sample and the white dwarfs analyzed by \citet{coutu19}. They report an EW detection limit of $0.5\,\AA$ for their sample, well below J-PLUS detection limit, providing a proper reference sample to test $\ewca$ as a proxy for the presence of calcium absorption. The calcium abundance, $\log {\rm Ca/He}$, as a function of $T_{\rm eff}$ from the spectroscopic analysis of \citet{coutu19} is presented in Fig.~\ref{fig:cahe_teff}. The common white dwarfs cover a wide range of properties, with $5\,000 < T_{\rm eff} < 15\,000$ K and $-11 < \log {\rm Ca/He} < -7$ dex. Low abundance systems are only present at cool temperatures due to observational bias.

The measured $\ewca$ shows the expected behavior, with larger values for larger calcium abundances at a given effective temperature (Fig.~\ref{fig:cahe_teff}, {\it left panel}). The typical uncertainty in the measured $\ewca$ of these sources is $\sigma_{\rm EW} = 5.9$\ \AA, that translates to a 99\% confidence (i.e., $S/N > 2.3$) in the detection of calcium absorption for systems with $\ewca \gtrsim 13.5$~\AA. We tested this limit by computing $\ewca$ from the synthetic J-PLUS photometry performed in a set of theoretical models \citep{koester10,koester11} for pure-He atmospheres polluted by metals at different effective temperatures ($4\,000 \leq T_{\rm eff} \leq 10\,000$ K in $100$ K steps) and calcium abundances ($-11 \leq \log {\rm Ca/He} \leq -6$ dex in $0.2$ dex steps). The models had a fixed surface gravity of $\log {\rm g} = 8$ dex and the abundance of other metals was set to that in the Solar System chondrites \citep{lodders03}. Then, we found at each $T_{\rm eff}$ the calcium abundance that produces $\ewca = 13.5$~\AA. The selection limit at temperatures larger than $T_{\rm eff} > 10\,000$~K was estimated from the values in \cite{coutu19}, properly scaled to match the selection limit from our models at the common temperature range, $8\,000 < T_{\rm eff} < 10\,000$ K. The combined selection curve agrees well with the 99$\%$ confidence detection for the data ({\it right panel} in Fig.~\ref{fig:cahe_teff}). We also found that $\ewca$ departs from the expectations at $T_{\rm eff} \lesssim 5\,500$ K, where the presence of metals greatly modifies the spectrum even at low abundances and the fitting to the J-PLUS photometry assuming unpolluted atmospheres is not valid anymore. Thus, we set $T_{\rm eff} = 5\,500$ K as the lower temperature limit in our study.

We conclude that the measured $\ewca$ provides a proxy for the presence of calcium absorption in metal-polluted white dwarfs, showing the expected trends with $T_{\rm eff}$ and Ca/He.

\subsection{Comparison between photometric and spectroscopic effective temperatures}\label{sec:teff_spec}
We aim to estimate the dependence in the fraction of calcium white dwarfs with the effective temperature. The temperatures from J-PLUS photometry assuming no metal pollution are compared with the values from \citet{coutu19}, as presented in Fig.~\ref{fig:teffspec}. We found an overestimation from J-PLUS of $4$\%, with no dependence on the effective temperature in the range $5\,000 < T_{\rm eff} < 15\,000$ K. This suggests a systematic difference due to our simplified models that assume white dwarfs atmospheres without metals. For consistency, we used the photometric effective temperatures from J-PLUS in the study of the calcium white dwarf population (Sect.~\ref{sec:fca}).

\subsection{EW$_{J0395}$ distribution of metal-polluted white dwarfs}\label{sec:caew_dist}
In this section, we present the distribution of $\ewca$ for the general white dwarf population and for those systems classified as metal-polluted in the GF21 catalog. Prior to this analysis, we selected white dwarfs with $5\,500 \leq T_{\rm eff} < 30\,000$ K to ensure that $\ewca$ is a proper proxy for the calcium absorption (Sect.~\ref{sec:cahe_teff_spec}) and the uncertainty in the measured effective temperature is below the $5$\% level \citep{clsj22pda}. We also selected objects with $M > 0.45\ M_{\odot}$ to discard low-mass systems from binary evolution and unresolved double degenerates. The final sample comprises $4\,399$ white dwarfs.

The $\ewca$ as a function of $T_{\rm eff}$ for the final sample is presented in Fig.~\ref{fig:ewdist}. The general distribution of the population is well described by a Gaussian with median $\mu = 0.5\,\AA$ and dispersion $\sigma = 8\,\AA$. The $33$ white dwarfs dominated by the presence of polluting metals (DZ, DZA, DZB, etc.), following GF21 classification, are located at $T_{\rm eff} < 15\,000$ K and $\ewca > 0~\AA$. Their distribution is asymmetric, with median $\overline{\rm EW}_{J0395} = 17~\AA$ and $16$\% to $84$\% of the distribution within $[7,53]$ \AA. A two-sized Kolmogorov-Smirnov test yields a probability lower than $10^{-14}$ that the general and the metal-dominated populations are drawn from the same parent sample.

The $21$ white dwarfs with a subdominant presence of metals (DAZ, DBZ, etc.) in the GF21 classification present lower values of $\ewca$ and cover a larger temperature range, reaching $T_{\rm eff} \sim 20\,000$ K. This is the expected trend, with a shallower calcium absorption than the DZ systems. The median and the [16\%,84\%] range are $\overline{\rm EW}_{J0395} = 4~\AA$ and $[1,10]$ \AA, respectively. In this case, the Kolmogorov-Smirnov test provides a probability of $0.006$, or a $2.5\sigma$ difference, between the general and the metal-polluted population.

The measured $\ewca$ has the expected behavior with the spectral type, presenting positive values for objects where the optical spectrum is dominated by metal lines and larger values for those classified as DZ. We note that $\ewca$ is statistically positive even for DAZ and DZB hybrid types. Interestingly, Figure~\ref{fig:ewdist} also shows that white dwarfs with significantly positive $\ewca$ and without SDSS spectroscopic counterpart are present in the J-PLUS + {\it Gaia} sample. 

As a summary, we found that the $\ewca$ and $\rm T_{\rm eff}$ derived from J-PLUS photometry and {\it Gaia} parallaxes are good proxies for the effective temperature and the Ca/He abundance derived from the spectra. Moreover, the $\ewca$ distribution of the metal-polluted systems presents a positive value and differs from the general population, dominated by non-polluted white dwarfs. The next step in our study was to model the observed $\ewca$ distribution, as detailed in the next section, to estimate the fraction of white dwarfs with calcium absorption in J-PLUS as a function of $T_{\rm eff}$.

\section{Evolution in the fraction of calcium white dwarfs with the effective temperature}\label{sec:fca}
In the previous section, we demonstrated that $\ewca$ is a good proxy to unveil the presence of calcium absorption in white dwarfs. However, the uncertainties in the measurements can blur the difference between genuine metal-polluted systems with $\ewca > 0~\AA$ and non-polluted white dwarfs with $\ewca = 0~\AA$. This limitation was solved with a Bayesian modeling of the observed $\ewca$ distribution that includes the effect of the uncertainties and permits to recover the real, underlying $\ewca$ distribution. The $4\,399$ white dwarfs with $30\,000 > T_{\rm eff} > 5\,500$ K and $M > 0.45\ M_{\odot}$ were used in the following analysis.

\subsection{Modeling of the distribution}
The real distribution of $\ewca$ in absence of observational uncertainties is
\begin{equation}
    D\,(\ewca^{\rm real}\,|\,\Theta),\label{eq:Dreal}
\end{equation}
where $\Theta$ is the set of parameters that define the distribution. The function was normalized to have a unity integral. To simplify the notation, we omit the normalization constants in the following.

The real values of $\ewca$ are blurred during the measuring process due to the observational uncertainties. Thus, the probability of observing an equivalent width for a given set of parameters is
\begin{equation}
\begin{split}
    P\,&(\ewca\,|\,\Theta,\sigma_{\rm EW}) = \\
    &\int  D\,(\ewca^{\rm real}\,|\,\Theta)\ P_{\rm G}\,(\ewca\,|\,\ewca^{\rm real},\sigma_{\rm EW})\ {\rm d}\ewca^{\rm real},\label{eq:dgauss}
\end{split}
\end{equation}
where Gaussian uncertainties were assumed. Combining the individual probabilities for the white dwarf population, the likelihood of the parameters is obtained as
\begin{equation}
    \mathcal{L}\,(\mathbf{EW_{J0395}}\,|\,\Theta,\boldsymbol{\sigma}_{\mathbf{EW}}) = \prod_i P\,(\ewca^i\,|\,\Theta,\sigma^i_{\rm EW}),
\end{equation}
where the index $i$ runs over the white dwarfs in the sample, and $\mathbf{EW_{J0395}}$ and $\boldsymbol{\sigma}_{\mathbf{EW}}$ are the vectors that comprise the individual measurements and their uncertainties, respectively. The posterior probability of the parameters was obtained by applying the priors,
\begin{equation}
    {\rm PDF}\,(\Theta) \propto \mathcal{L}\,(\mathbf{EW_{J0395}}\,|\,\Theta,\boldsymbol{\sigma}_{\mathbf{EW}}) \ P\,(\Theta).
\end{equation}

We explored the parameters space with the \texttt{emcee} code \citep{emcee}, a \texttt{Python} implementation of the affine-invariant ensemble sampler for the Markov chain Monte Carlo (MCMC) technique proposed by \citet{goodman10}. The \texttt{emcee} code provides a collection of solutions, noted $\Theta_{\rm MC}$, with their density being proportional to the posterior probability of the parameters. We obtained the central values of the parameters and their uncertainties from a Gaussian fit to the distribution of the solutions, noted $\overline{\Theta}_{\rm MC}$ and $\sigma_{\Theta}$, respectively.

To compare the different models, we used the Bayesian information criterion (BIC, \citealt{schwarz78}), defined as
\begin{equation}
{\rm BIC} = N\log n - 2\log {\mathcal L}\,(\mathbf{EW_{J0395}}\,|\,\overline{\Theta}_{\rm MC}, \boldsymbol{\sigma}_{\mathbf{EW}}),
\end{equation}
where $N$ is the number of parameters in the model and $n$ the number of white dwarfs in the sample. A BIC difference between models larger than $10$ is a strong support for the model with the lower BIC.

\subsection{Temperature-independent models}
Models that do not depend on $T_{\rm eff}$ were explored first. This provided us with an initial guess about the relevance of the metal-polluted population. The results for the different models are summarized in Table~\ref{tab:models}.

The baseline model assumes a unique population of white dwarfs without metal pollution. This is described with a Dirac delta function,
\begin{equation}
    D_0 = \delta\,(\ewca).
\end{equation}
This model has no free parameters and its likelihood was used to normalize all the models in the following. This implies $\log \mathcal{L}_0~=~0$ and ${\rm BIC} = 0$. We note that the notation was simplified by omitting the dependent variable $\ewca^{\rm real}$ and only the parameters of the distribution were explicit. 

\begin{table} 
\caption{Results of the Bayesian fitting to the observed $\ewca$ distribution.}
\label{tab:models}
\centering 
        \begin{tabular}{l c c c c}
        \hline\hline\rule{0pt}{3ex} 
        Model   &  Fitting results & $\log \mathcal{L}$ & BIC & $\Delta$ BIC \\
                & $\overline{\Theta}_{\rm MC} \pm \sigma_{\Theta}$ & & & \\
        \hline\rule{0pt}{3ex}
        \!$D_0$            & $\cdots$   &  $0$  &  $0$  &    \\ 
        \rule{0pt}{3ex} $D_1$            & $f_{\rm Ca} = 0.016 \pm 0.007$            &  $1815$  &  $-3622$  &  $-3622$   \\ 
       \rule{0pt}{3ex}  $D_1^{\prime}$   & $f_{\rm Ca}^{\prime} = 0.002 \pm 0.002$   &  $0$ &   $8$  &  $8$  \\ 
       \rule{0pt}{5ex}  $D_2$            & \begin{tabular}{@{}c@{}} $f_{\rm Ca} = 0.035 \pm 0.007$ \\ $\alpha = -0.042 \pm 0.007$\end{tabular}   & $1832$  & $-3647$ & $-25$ \\ 
       \rule{0pt}{6ex}  $D_3$            & \begin{tabular}{@{}c@{}} $a = 0.18 \pm 0.03$ \\ $b = 0.067 \pm 0.011$ \\ $\alpha = -0.047 \pm 0.007$\end{tabular}   & $1872$  & $-3719$ & $-72$\\
        \hline 
\end{tabular}
\tablefoot{The last column displays the difference in the Bayesian information criteria (BIC) between consecutive models. Negative values favor the model with lower BIC.
}
\end{table}

\begin{figure*}[t]
\centering
\resizebox{0.49\hsize}{!}{\includegraphics{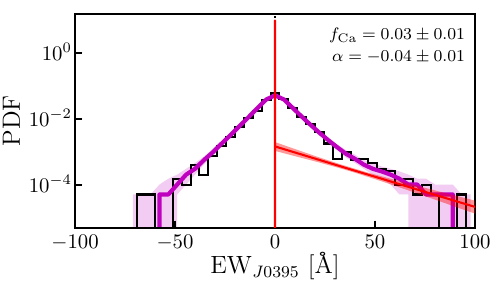}}
\resizebox{0.49\hsize}{!}{\includegraphics{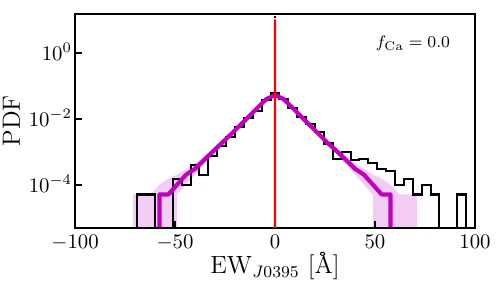}}
\caption{Probability distribution function of the observed $\ewca$ (black histogram). The {\it left panel} shows the best-fit $D_2$ model and its 68\% confidence interval with the red line and area, respectively. The best-fit parameters are labeled in the panel. The purple line and area show the median distribution and its 68\% confidence interval obtained from $5\,000$ random samples extracted from $D_2$ and affected by the same uncertainties $\sigma_{\rm EW}$ than the observations. The {\it right panel} is similar to the {\it left panel} but for the baseline model $D_0$.}
\label{fig:ewca_model}
\end{figure*}

The next step was to include a second population with positive values of $\ewca$. Our first attempt was to assume a constant density,
\begin{equation}
    D_1\,(f_{\rm Ca}) = (1 - f_{\rm Ca}\,)\,D_0 + f_{\rm Ca}\,H(\ewca),
\end{equation}
where $f_{\rm Ca}$ is the total fraction of white dwarfs with calcium absorption and $H$ is the Heaviside step function. A prior in $f_{\rm Ca}$ was applied to restrict its value between zero and one. This model is overwhelmingly favored by the data, with ${\rm BIC} = -3\,622$. The obtained fraction of calcium white dwarfs was $f_{\rm Ca} = 0.016 \pm 0.002$. We discarded that this positive signal is due to systematics in the measurements or in the uncertainties by testing the presence of nonphysical calcium emitters,
\begin{equation}
    D_1^{\prime}\,(f_{\rm Ca}^{\prime}) = (1 - f_{\rm Ca}^{\prime}\,)\,D_0 + f_{\rm Ca}^{\prime}\,H(-\ewca).
\end{equation}
We found $\log \mathcal{L}_1^{\prime} = 0$ and a fraction of calcium emitters that tend to zero. Because the model has one parameter, we obtain ${\rm BIC} = 8$ and the baseline model is favored. Hence, we found that the $\ewca$ distribution presents a relevant population of metal-polluted white dwarfs with calcium absorption.

Then, a dependence on equivalent width was included as
\begin{equation}
    D_2\,(f_{\rm Ca},\alpha) = (1 - f_{\rm Ca}\,)\,D_0 + f_{\rm Ca}\,D_{\rm Ca}\,(\alpha),
\end{equation}
where 
\begin{equation}
    D_{\rm Ca}\,(\alpha) = {\rm exp}(\alpha\ \ewca)\,H(\ewca).\label{eq:Dca}
\end{equation}
We found that the data favor a decrease in the number of calcium white dwarfs with equivalent width, $\alpha = -0.042 \pm 0.007$ and $\Delta {\rm BIC} = -25$ with respect to the model $D_1$. The total fraction of white dwarfs with calcium absorption in this model was $f_{\rm Ca} = 0.035 \pm 0.007$.

A key aspect of the performed analysis is the accounting for the impact of the observational errors in the broadening of the distributions. We highlight this point by comparing the observed $\ewca$ distribution and the best $D_2$ distribution in Fig.~\ref{fig:ewca_model}. At first glance, both distributions seem to differ significantly. Nevertheless, the model must be convolved with the same observational uncertainties in $\ewca$ for a proper comparison with the observed distribution. Thus, we extracted $5\,000$ random samples of $n = 4\,399$ sources from $D_2$, which correspond to $\ewca^{\rm real}$ in Eq.~(\ref{eq:Dreal}). Then, we randomly assigned to them the $\sigma_{\rm EW}$ from the observations, and drawn a random value from a Gaussian with median $\ewca^{\rm real}$ and dispersion $\sigma_{\rm EW}$. Finally, the histograms of the random samples were estimated to obtain the median simulated distribution and its uncertainty. The comparison between the simulated and the observational distributions is excellent (Fig.~\ref{fig:ewca_model}), demonstrating the goodness of the fitting process and the impact of the errors in the underlying, real distribution of $\ewca$.

As a final test, we repeated the procedure above with the baseline model $D_0$, that is, assuming that there ire no objects with calcium absorption in the sample. In this case the negative range is recovered, but the positive range presents a clear excess of objects with large equivalent widths. This visually supports the BIC results summarized in Table~\ref{tab:models}.

\begin{figure}[t]
\centering
\resizebox{\hsize}{!}{\includegraphics{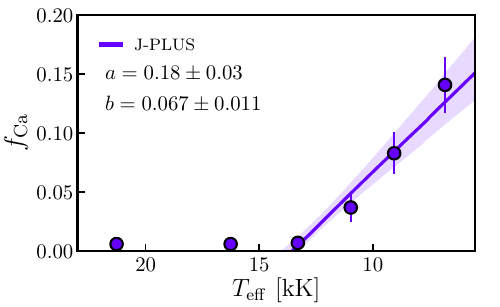}}
\caption{Fraction of white dwarfs with calcium absorption ($f_{\rm Ca}$) as a function of the effective temperature ($T_{\rm eff}$). The best-fit $F_{\rm Ca}$ model and its 68\% confidence interval are shown with the purple line and area, respectively. The parameters of the linear trend from Eq.~(\ref{eq:Fca}) are presented in the panel. The purple dots depict the fractions obtained with the temperature-independent model $D_2$ at different $T_{\rm eff}$ intervals.
}
\label{fig:fca_teff_jplus}
\end{figure}

\subsection{Temperature-dependent model}\label{sec:fca_teff}
The final step in the modeling of the $\ewca$ distribution was to include a dependence on the effective temperature for the fraction of calcium white dwarfs,
\begin{equation}
    \begin{split}
    D_3\,(T_{\rm eff},a&,b,\alpha) = \\
    & [1 - F_{\rm Ca}(T_{\rm eff},a,b)]\,D_0 + F_{\rm Ca}(T_{\rm eff},a,b)\,D_{\rm Ca}\,(\alpha),   
    \end{split}
\end{equation}
where 
\begin{equation}
    F_{\rm Ca}(T_{\rm eff},a,b) = b - a\times \bigg( \frac{T_{\rm eff}}{10^4\ {\rm K}} - 1 \bigg).\label{eq:Fca}
\end{equation}
The model $D_3$ has three free parameters, as the effective temperatures of the white dwarfs were measured. We assumed $T_{\rm eff}$ without error in the fitting process.

The main result of the present paper is the change in the fraction of white dwarfs with calcium absorption along the cooling sequence from Eq.~(\ref{eq:Fca}). We found $b = 0.067 \pm 0.011$ and $a = 0.18 \pm 0.03$, as presented in Fig.~\ref{fig:fca_teff_jplus}. The significance in the slope parameter $a$ is at 6$\sigma$ level, with a larger fraction of calcium white dwarfs at lower temperatures. The BIC difference with respect to the best temperature-independent model is $\Delta \rm{BIC} = -72$, implying that the evolution in the fraction of calcium white dwarfs with the effective temperature is largely supported by the data. 

We also tried a model with temperature variation in the $\alpha$ parameter. The obtained BIC difference with the model $D_3$ was $3$ and there is therefore no evidence for a significant temperature variation in the parameter $\alpha$. The best-fit value from model $D_3$ was $\alpha = -0.047 \pm 0.007$.

We complemented the result by computing $f_{\rm Ca}$ at different temperature ranges with model $D_2$ and fixing $\alpha = -0.047$. The derived fractions are summarized in Table~\ref{tab:fca_teff} and shown in Fig.~\ref{fig:fca_teff_jplus}. The fractions from the individual temperature ranges are compatible with the inference from the global analysis.

\begin{figure*}[t]
\centering
\resizebox{0.49\hsize}{!}{\includegraphics{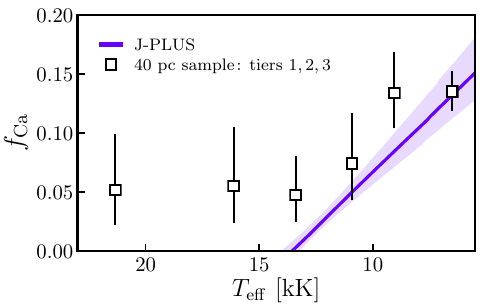}}
\resizebox{0.49\hsize}{!}{\includegraphics{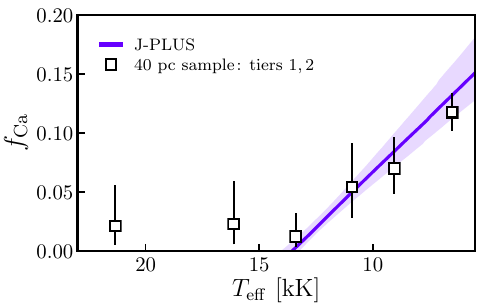}}\\
\resizebox{0.49\hsize}{!}{\includegraphics{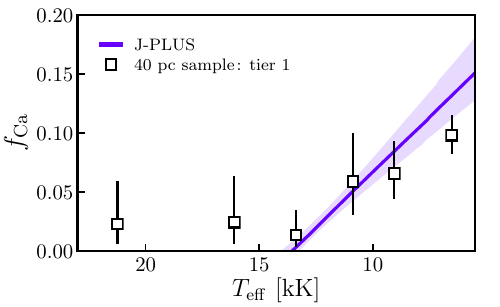}}
\resizebox{0.49\hsize}{!}{\includegraphics{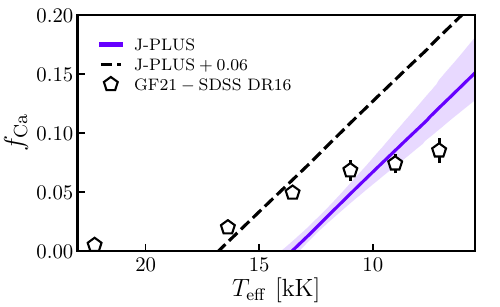}}
\caption{Fraction of white dwarfs with calcium absorption ($f_{\rm Ca}$) as a function of the effective temperature ($T_{\rm eff}$). The J-PLUS model and its 68\% confidence interval are shown with the purple line and area, respectively. The squares show the fractions derived from the 40 pc sample for three different tiers with increasing sensitivity to the presence of metal lines (see text for details). The pentagons are the fractions estimated with the SDSS DR16 spectra gathered in the GF21 catalog. The dashed line depicts the J-PLUS model with an addition of 0.06.}
\label{fig:fca_teff}
\end{figure*}

\begin{table*} 
\caption{Fraction of calcium white dwarfs measured at different effective temperature intervals.}
\label{tab:fca_teff}
\centering 
        \begin{tabular}{@{\extracolsep{3pt}}ccccccccccc@{}}  
        \hline\hline\noalign{\smallskip}            
        &   &    \multicolumn{2}{c}{J-PLUS}   &    \multicolumn{4}{c}{40 pc spectroscopic sample}  &    \multicolumn{3}{c}{SDSS spectroscopy}\\
        \noalign{\smallskip}\cline{3-4}\cline{5-8}\cline{9-11}\noalign{\smallskip}
        $T_{\rm eff}^{\rm min}$   & $T_{\rm eff}^{\rm max}$ & $\langle T_{\rm eff} \rangle$ & $f_{\rm Ca}$ & $\langle T_{\rm eff} \rangle$ & $f_{\rm Ca}$ &  $f_{\rm Ca}$  & $f_{\rm Ca}$ & $S/N_{\rm min}$ & $\langle T_{\rm eff} \rangle$ & $f_{\rm Ca}$\\\rule{0pt}{3ex}
                [kK]        &                [kK] &                   [kK]    &     &       [kK]    &   Tier 1,2,3      &    Tier 1,2       &   Tier 1  & & [kK] & \\
        \hline\rule{0pt}{3ex}

        $5.5$     &   $8$  & $6.8$ & $0.14 \pm 0.02$ & $6.5$ & $0.13 \pm 0.02$        & $0.12 \pm 0.02$  &  $0.10 \pm 0.02$   &   $15$ &  $7.1$  &  $0.085\pm0.011$ \\\rule{0pt}{3ex}

        $8$      &   $10$  & $9.1$ & $0.08 \pm 0.02$ & $9.0$ & $0.13 \pm 0.03$        &  $0.07^{+0.03}_{-0.02}$  & $0.07^{+0.03}_{-0.02}$ &   $14$ &  $9.0$  &  $0.074\pm0.008$ \\\rule{0pt}{3ex}

        $10$     &   $12$  & $11.0$ & $0.04 \pm 0.01$ & $10.9$ & $0.07^{+0.04}_{-0.03}$  & $0.05^{+0.04}_{-0.03}$  & $0.06^{+0.04}_{-0.03}$ &   $13$ &  $11.0$  &  $0.068\pm0.009$ \\\rule{0pt}{3ex}

        $12$     &   $15$  & $13.3$ & $0.01 \pm 0.01$ & $14.4$ & $0.05^{+0.03}_{-0.02}$ &  $0.01^{+0.02}_{-0.01}$ & $0.01^{+0.02}_{-0.01}$ &   $10$ &  $13.5$  &  $0.049\pm0.005$ \\\rule{0pt}{3ex}

        $15$     &   $18$  & $16.3$ & $0.01 \pm 0.01$ & $16.1$ & $0.06^{+0.05}_{-0.03}$  & $0.02^{+0.04}_{-0.02}$   & $0.02^{+0.04}_{-0.02}$ &   $6$ &  $16.4$  &  $0.020\pm0.003$ \\\rule{0pt}{3ex}

        $18$     &   $30$  & $21.3$ & $0.01 \pm 0.01$ & $21.2$ & $0.05^{+0.05}_{-0.03}$ & $0.02^{+0.03}_{-0.02}$  & $0.02^{+0.04}_{-0.02}$ &   $5$ &  $22.2$  &  $0.005\pm0.001$ \rule{0pt}{3ex}\\
        \hline 
\end{tabular}
\end{table*}

The trend in Fig.~\ref{fig:fca_teff_jplus} should be understood as a reflection of the selection effects discussed in Sects.\ref{sec:intro} and \ref{sec:ewca_test}. Assuming a total fraction of metal-polluted white dwarfs of $30$\%-$50$\% \citep{zuckerman03,zuckerman10,koester14} without any effective temperature dependence, the change in the equivalent width detection limit with $T_{\rm eff}$ at a constant calcium abundance (Fig.~\ref{fig:cahe_teff}) naturally creates the observed trend. We stress that the $f_{\rm Ca}$ obtained from the modeling of the $\ewca$ distribution is not equivalent to the fraction of white dwarfs above a given equivalent width, but proportional due to the functional form assumed to describe the population of white dwarfs with calcium absorption in Eq.~(\ref{eq:Dca}). 

The key point of our result is that the observed $f_{\rm Ca}$ encodes the selection function of calcium white dwarfs in the optical from the total population of metal-polluted systems. Hence, the comparison of the J-PLUS results with the fraction of calcium white dwarfs obtained from spectroscopic samples permits to better understand the impact of the target selection in the spectroscopic case, as explored in the following Section.

\section{Selection function in the detection of metal-polluted white dwarfs}\label{sec:selec_spec}
We complement the photometric results from J-PLUS in the previous section with the spectroscopic values from the 40 pc sample (Sect.~\ref{sec:40pc}), derived from heterogeneous spectroscopy with a variety of sensitivities, and the SDSS DR16 spectroscopic sample (Sect.~\ref{sec:sdss}), that presents an homogeneous sensitivity. The three samples under study originated from the GF21 catalog and were selected by distance (40 pc sample), magnitude ($r < 19.5$ mag, J-PLUS), and colors (SDSS). Thus, we can combine the three datasets to learn about the selection of metal-polluted white dwarfs in the optical and better understand the impact of color selections on the SDSS spectroscopic sample. 

\subsection{J-PLUS and the 40 pc sample}\label{sec:40pc}
The results from the 40 pc sample are presented in Fig.~\ref{fig:fca_teff} and Table~\ref{tab:fca_teff}. We used the effective temperature and the mass derived using {\it Gaia} photometry and astrometry by GF21 assuming pure-H models for DA as principal type, and pure-He models at $T_{\rm eff} < 6\,600$ K and mixed models with $\log {\rm H/He} = -5$ otherwise for non-DAs. These choices were similar in the case of J-PLUS (Sect.~\ref{sec:methods}), minimizing systematic differences in $T_{\rm eff}$ and mass between both data sets. The same $T_{\rm eff}$ bins and mass limit $M > 0.45\, M_{\odot}$ as in the J-PLUS analysis were used to compute the fraction of white dwarfs with signs of metal pollution (DZ, DZA, DZB, DAZ, DBZ, and other hybrid types). Because of the low number statistics, the uncertainties were computed using the Bayesian binomial recipe in \cite{cameron11}. We find $f_{\rm Ca} \approx 0.05$ at $T_{\rm eff} > 13\,500$ K, an increase in the fraction up to $f_{\rm Ca} \approx 0.13$ at $T_{\rm eff} \sim 8\,000$ K, and then a new plateau at lower temperatures. The fraction from the 40 pc sample is larger than in J-PLUS except at the lower temperature bin.

As discussed in \citet{40pciii}, the detectability of metals in the spectra has a dependence with the resolution of the data: the larger the resolution, the lower the equivalent width of the calcium (or any other metal) absorption than can be detected. Taking this into account, a mismatch between J-PLUS and the 40 pc sample is not surprising, and an offset is expected due to the different selection limits of both data sets. In this regard, an offset of $0.04$ applied to the J-PLUS results reconciles the measurements at $T_{\rm eff} > 8\,000$ K.

However, the different sensitivity with the spectral resolution also affects the 40 pc sample, that gathers information from several instruments and works obtained with a diversity of configurations. We defined three tiers based on the resolution and the sensitivity to detect the presence of metal lines: tier $1$ is composed by those data with $R \sim 500 - 2\,000$; tier $2$ by the X-Shooter spectra, with a resolution of $R = 5\,400$ in the blue; and tier $3$ by those classifications based on $R > 18\,500$ spectra and with a measured calcium equivalent width lower than $0.5$ $\AA$. When tier $3$ objects are assumed as white dwarfs without metal pollution, this is, calcium should not be detected with the tier $1$ and $2$ configurations, we obtained a lower $f_{\rm Ca}$ at all temperatures ({\it top right panel} in Fig.~\ref{fig:fca_teff}). Indeed, all the metal-polluted objects at $T_{\rm eff} > 12\,000$ K belong to tier $3$ and $f_{\rm Ca} \sim 0$ at the high-temperature end, in agreement with the J-PLUS results. The fractions at lower temperatures also agree with the J-PLUS model. As stated before, the match between both works should be considered circumstantial and the relevant result is the consistent increase in the fraction of calcium white dwarfs found at $T_{\rm eff} < 13\,500$ K by both data sets.

Finally, we also remove from the metal-polluted sample the white dwarfs in the tier $2$ observed by X-Shooter ({\it bottom left panel} in Fig.~\ref{fig:fca_teff}). Following the previous discussion, the fraction of metal-polluted objects decreases in the lower temperature bins, the range where the X-Shooter spectra are concentrated. The trends are similar to the previous case, and therefore, the conclusions are the same.

The fraction of calcium white dwarfs based on spectroscopy has a dependence on the instrumental setup of the observations. Nevertheless, the volume-limited 40 pc sample with heterogeneous spectroscopic information provides a consistent picture with the photometric results from J-PLUS, showing an increase in $f_{\rm Ca}$ at $T_{\rm eff} < 13\,500$ K.

\subsection{J-PLUS and the SDSS spectroscopic sample}\label{sec:sdss}
We repeated the estimation in the fraction of calcium white dwarfs using the SDSS DR16 spectroscopic information included in the GF21 catalog. As in the 40 pc sample, the atmospheric parameters from GF21 were used, and the same effective temperature and mass ranges than in J-PLUS were assumed. In this case all the SDSS DR16 data have a similar resolution. We minimized the impact of the different EW limit for spectra with different $S/N$ by using only sources with $S/N > S/N_{\rm min}$ at each temperature bin. We choose $S/N_{\rm min}$ to have a mean $S/N \approx 22.5$ in all the considered bins. We also tested a common limit for all the sources and the results were similar. We summarized the obtained fraction of metal-polluted white dwarfs from SDSS DR16 spectroscopy in the {\it bottom right panel} of Fig.~\ref{fig:fca_teff} and Table~\ref{tab:fca_teff}.

Compared with J-PLUS, the SDSS-based fractions are larger at $T_{\rm eff} > 10\,000$ K and smaller at lower effective temperatures. The trend is not consistent with the results from J-PLUS and the 40 pc sample, with an increase below $T_{\rm eff} \sim 20\,000$ K and a plateau of $f_{\rm Ca} \approx 0.075$ at $T_{\rm eff} \lesssim 12\,000$ K. The SDSS spectra share a similar resolution and $S/N$, thus the main difference with respect to J-PLUS and the 40 pc sample is the additional color selection of SDSS spectroscopy. We recall that the three samples under consideration were initially selected from the GF21 catalog.

As discussed in the previous section, an offset between the spectroscopy and J-PLUS is expected due to the different sensitivities. We can match J-PLUS and SDSS-based fractions at $T_{\rm eff} > 12\,000$ K by including a $0.06$ increase to the J-PLUS fitting. However, the extrapolation of the J-PLUS trend to lower temperatures implies that SDSS spectra are selected against cool metal-polluted white dwarfs and the incompleteness at $T_{\rm eff} \sim 7\,000$ K could reach $50$\%.

\begin{figure}[t]
\centering
\resizebox{\hsize}{!}{\includegraphics{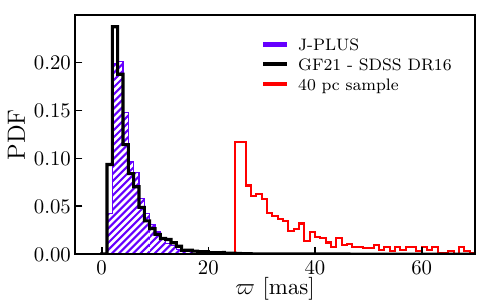}}
\caption{Probability distribution function of the parallax ($\varpi$) for the J-PLUS sample (purple), the 40 pc sample (red) and the SDSS DR16 spectroscopic sample gathered in the GF21 catalog (black).}
\label{fig:pxdist}
\end{figure}

The SDSS selection against cool, metal-polluted white dwarfs was already pointed out by \citet{dufour07} and \citet{hollands17}. The SDSS programs for white dwarfs target the color space where non-polluted objects are expected \citep[e.g.][]{kepler16}. Metal-polluted white dwarfs tend to have redder $(u-g)$ colors and the cooler ones have been discovered in SDSS spectra targeting quasars at redshifts larger than $2.5$ \citep{hollands17}, far from the expected locus of non-polluted white dwarfs. Our results support this scenario.

\subsection{Summary and future prospects}
We found an increase in the fraction of white dwarfs with \ion{Ca}{ii} H+K absorption along the cooling sequence, growing from $f_{\rm Ca} \approx 0$ at $T_{\rm eff} = 13\,500$ K to $f_{\rm Ca} \approx 0.15$ at $T_{\rm eff} = 5\,500$ K. This trend is observed in J-PLUS photometric data and in the 40 pc spectroscopic sample, both selected from the GF21 catalog. The SDSS DR16 spectra of the GF21 catalog present a deficit of metal-polluted white dwarfs at $T_{\rm eff} < 12\,000$ K. We conclude that this is due to the color selections affecting SDSS spectroscopy.

We discarded that the observed trends are dictated by the parallax (distance) distribution of the samples, as presented in Fig.~\ref{fig:pxdist}. On the one hand, the 40 pc has $\varpi > 25$ mas by definition. On the other hand, the J-PLUS and SDSS samples present similar distributions with a peak at $\varpi \approx 3$ mas and a steady decrease at larger parallaxes. Despite the evident difference in the distance distribution, J-PLUS and the 40 pc provide compatible results. This reinforces the color selection in SDSS spectroscopy as the source of the observed discrepancy.

As a consequence, the accretion rates and metal abundances of the white dwarf population based on SDSS spectroscopy should be affected by the aforementioned selection effect. The on-going and future observations from the SDSS-V Milky Way mapper \citep{sdssv}, the William Herschel Telescope Enhanced Area Velocity Explorer (WEAVE, \citealt{weave}), the Dark Energy Spectroscopic Instrument (DESI, \citealt{desi_mws}), and the 4-metre Multi-Object Spectrograph Telescope (4MOST, \citealt{4most_wd}), that plan to follow up one hundred thousand white dwarfs from the GF21 catalog, will provide homogeneous spectroscopic data sets with a well-defined selection function to overcome the current SDSS limitations. In addition, these spectroscopic samples will test the trend found by both J-PLUS and the 40 pc sample.

Recently, the early data release (EDR) of DESI presented spectra for $2\,706$ white dwarfs \citep{desi_edr_wd}. This permitted to analyze the metal-polluted fraction for objects with helium-dominated atmospheres, finding a nearly constant value of $0.21 \pm 0.03$ along the cooling sequence \citep{desi_edr_fca}. The comparison with DESI is beyond the scope of the present paper, and will be explored in the future.

\section{Probability of presenting calcium absorption}\label{sec:pca}
As an application of our results, we estimate the probability of each individual white dwarf to have calcium absorption. We assumed that a white dwarf can be in two states, $s = \{0,1\}$, corresponding to the absence (state 0) or the presence (state 1) of calcium absorption. Using the results from the model $D_3$ presented in Sect.~\ref{sec:fca_teff}, the probability of having calcium absorption was defined as
\begin{equation}
    p_{\rm Ca} = \frac{P\,(1)\,P\,(\ewca\,|\,1,\sigma_{\rm EW})}{P\,(0)\,P\,(\ewca\,|\,0,\sigma_{\rm EW}) + P\,(1)\,P\,(\ewca\,|\,1,\sigma_{\rm EW})},
\end{equation}
where $P\,(1) = F_{\rm Ca}(T_{\rm eff},a,b)$ is the prior for having calcium absorption, $P\,(0) = 1 - P\,(1)$, and $P\,(\ewca\,|\,s,\sigma_{\rm EW})$ was estimated with Eq.~(\ref{eq:dgauss}) using $D = D_0$ for $s = 0$ and $D = D_{\rm Ca}$ for $s = 1$.

The obtained $\pca$ has two practical applications: can be used to properly weight and study the properties of the calcium white dwarf population (Sect.~\ref{sec:pca_spec}) or to provide high-confidence candidates of metal-polluted white dwarfs for spectroscopic follow up (Sect.~\ref{sec:pca_gtc}). These two applications also help to test the reliability of the results presented in Sect.~\ref{sec:fca_teff}.

\begin{figure}[t]
\centering
\resizebox{\hsize}{!}{\includegraphics{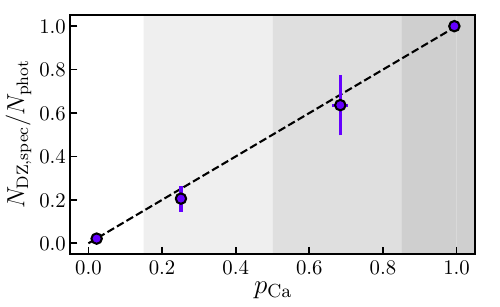}}
\caption{Fraction of white dwarfs with metal-pollution from spectroscopy as a function of the probability of having calcium absorption from J-PLUS photometry, $\pca$ (purple dots). The gray areas show the four intervals in $\pca$ analyzed. The dashed line marks the one-to-one relation.}
\label{fig:pca}
\end{figure}

\begin{table*} 
\caption{White dwarfs in J-PLUS DR2 with a probability of having calcium absorption $p_{\rm Ca} > 0.99$.}
\label{tab:dzcand}
\centering 
        \begin{tabular}{l c c c c c}  
        \hline\hline\rule{0pt}{3ex} 
        Tile - number   & RA & Dec & EW$_{J0395}$ & Spectroscopic type & Reference \\
                        & [deg] & [deg] & [$\AA$] &                  \\
        \hline\rule{0pt}{3ex}
\!62448-492 &109.5152   &24.1653    & $49 \pm 7$    & $\cdots$      & $\cdots$ \\    
62877-17245 &140.6428   &30.6357    & $63 \pm 8$    & $\cdots$      & $\cdots$ \\    
63200-15839 &132.6463   &32.1344    & $20 \pm 3$    &DABZ           &1 \\            
63260-7923  &141.3462   &31.5048    & $93 \pm 14$   &DZ             &2 \\            
64175-5622  &131.6842   &35.6424    & $55 \pm 1$    &DZA            &3 \\            
64246-13960 &211.0440   &36.3488    & $78 \pm 8$    &DZ             &2 \\            
64391-23916 &121.5123   &37.7889    & $49 \pm 9$    &DZ             &4 \\            
65004-14325 &155.7131   &39.0699    & $32 \pm 5$    &DZ             &5 \\            
65191-37967 &275.1556   &39.0642    & $19 \pm 5$    & $\cdots$      & $\cdots$ \\    
65287-10021 &110.2899   &39.9275    & $56 \pm 5$    &DZ             &6 \\            
65538-13754 &147.8326   &40.5561    & $22 \pm 3$    &DZ             &4 \\            
65573-5008  &152.5324   &39.8144    & $55 \pm 4$    &DZA            &7 \\            
65703-11332 &179.9955   &40.7614    & $49 \pm 8$    &DZ             &8 \\            
66232-18038 &153.9924   &41.6920    & $46 \pm 7$    &DZA            &7 \\            
66690-7017  &232.8718   &42.6710    & $41 \pm 4$    &DZ             &4 \\            
66878-54289 &280.1593   &43.5647    & $47 \pm 3$    &DZA            &This work \\    
67798-3868  &231.3645   &48.1478    & $58 \pm 8$    &DZ             &This work \\    
67870-21693 &277.6252   &48.6408    & $67 \pm 15$   & $\cdots$      & $\cdots$ \\    
67940-5203  &225.5978   &49.5532    & $48 \pm 8$    &DZ             &9 \\            
68119-45138 &284.6112   &50.2898    & $23 \pm 3$    &DZ             &This work \\    
68360-9741  &136.4838   &52.5925    & $71 \pm 6$    &DZ             &10 \\           
68984-28039 &208.0738   &55.3940    & $75 \pm 6$    & $\cdots$      & $\cdots$  \\   
69864-7931  &190.8573   &60.8745    & $50 \pm 10$   & $\cdots$      & $\cdots$ \\    
70349-1777  &155.3859   &67.4214    & $57 \pm 3$    &DZ             &This work \\    
70362-19530 &156.3670   &68.7340    & $44 \pm 9$    & $\cdots$      & $\cdots$ \\    
71303-24934 &261.5943   &79.6338    & $43 \pm 3$    & $\cdots$      & $\cdots$ \\    
71463-9516  &  1.4895   & 0.3092    & $37 \pm 6$    &DZ             &10  \\          
71774-945   & 25.7520   & 1.2322    & $55 \pm 13$   &DZ             &2  \\           
71823-6314  &  1.5127   & 2.9625    & $75 \pm 7$    &DZ             &This work  \\   
71837-6175  &  3.2641   & 3.0511    & $66 \pm 14$   & $\cdots$      & $\cdots$ \\    
72226-12752 & 22.1569   & 5.1495    & $66 \pm 12$   & $\cdots$      & $\cdots$ \\    
72766-4002  & 11.6858   & 7.0743    & $28 \pm 6$    &DZ             &11  \\          
73076-14122 &337.0782   &10.6686    & $64 \pm 7$    &DZ             &This work  \\   
73205-4997  &336.0876   &11.1704    & $42 \pm 6$    & $\cdots$      & $\cdots$ \\    
73205-23737 &337.0084   &12.1257    & $53 \pm 2$    &DZ             &4  \\           
73715-747   &345.0930   &22.0713    & $35 \pm 4$    &DZ             &12  \\          
74106-16189 &358.9366   &29.7721    & $41 \pm 8$    &DZ             &13\\            
74780-6928  & 31.8918   &33.5248    & $21 \pm 5$    & $\cdots$      & $\cdots$ \\    
74934-1784  & 37.2103   &36.0513    & $45 \pm 3$    & $\cdots$      & $\cdots$ \\    
\hline 
\end{tabular}
   \tablebib{
    (1)~\citet{kong19}; (2) \citet{koester11}; (3) \citet{giammichele12}; (4) \citet{eisenstein06_dr4}: (5) \citet{limoges15}; (6) \citet{kepler15}; (7) \citet{dufour07}; (8) \citet{kleinman13}; (9) \citet{kilic20}; (10) \citet{kleinman04}; (11) \citet{GF15}; (12) \citet{limoges13}; (13) \citet{carter13}.
    }
\end{table*}

\begin{figure*}[t]
\centering
\resizebox{0.49\hsize}{!}{\includegraphics{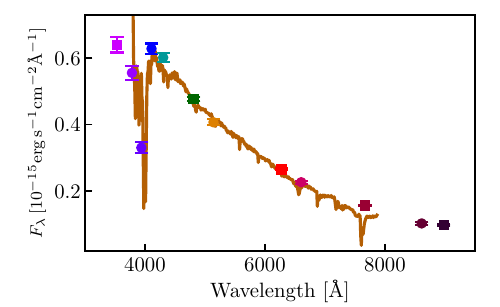}}
\resizebox{0.49\hsize}{!}{\includegraphics{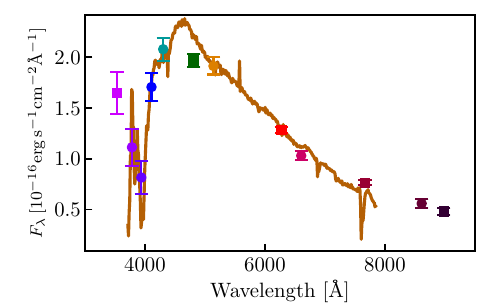}}\\
\resizebox{0.49\hsize}{!}{\includegraphics{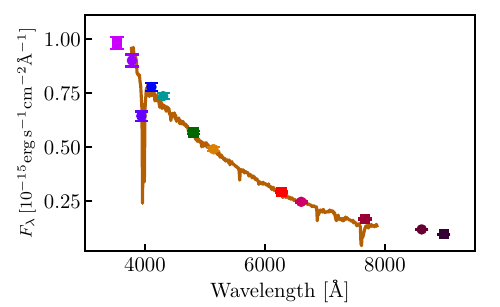}}
\resizebox{0.49\hsize}{!}{\includegraphics{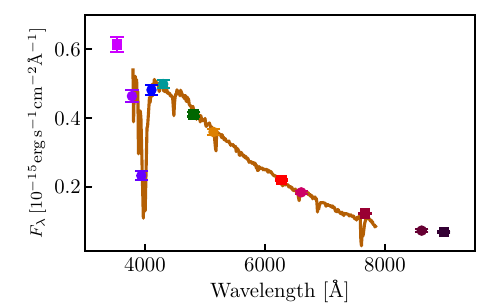}}\\
\resizebox{0.49\hsize}{!}{\includegraphics{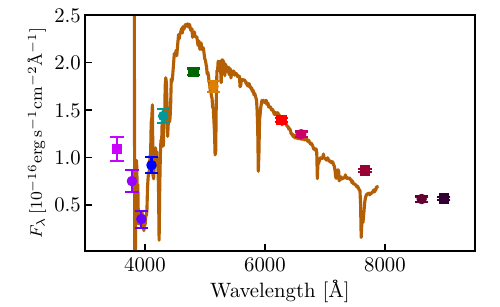}}
\resizebox{0.49\hsize}{!}{\includegraphics{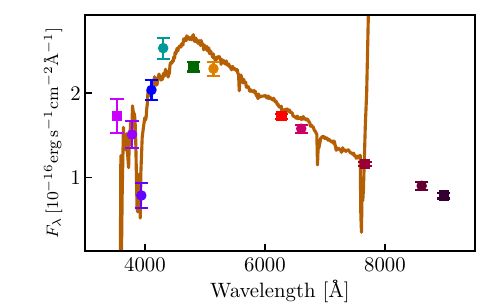}}
\caption{Six metal-polluted white dwarfs from the high-confidence J-PLUS sample confirmed with spectra from OSIRIS at GTC (brown lines). The symbols showing the J-PLUS photometry follow Fig.~\ref{fig:intro}. The spectra have been scaled to the $r$-band flux from J-PLUS.}
\label{fig:dzcand}
\end{figure*}

\subsection{Comparison with the spectroscopic classification}\label{sec:pca_spec}
The performance of a binary classification problem can be assessed with the purity, completeness, and other summary statistics for a given selection threshold in the probability. Nevertheless, we are interested in the reliability of the computed probabilities $p_{\rm Ca}$. It can be estimated by comparing the fraction of spectroscopic metal-polluted white dwarfs for a spectroscopic sample selected in a range of $p_{\rm Ca}$ against the median $p_{\rm Ca}$ in that range. A one-to-one relation translates to a well calibrated probability that provides the right fraction of true positives for a given $p_{\rm Ca}$.

We used the $1\,393$ white dwarfs with SDSS DR16 spectroscopic classification and $S/N > 5$ in the GF21 catalog of the sample with $M > 0.45\ M_{\odot}$ and $5\,500 < T_{\rm eff} < 30\,000$ K. We found $59$ metal-polluted systems and $1\,334$ white dwarfs without metal pollution, including DAs, DBs, DCs, DQs and hybrid types without evidence of metals. The fraction of spectroscopic white dwarfs with metals as a function of $\pca$ is presented in Fig.~\ref{fig:pca}. The number of metal-polluted objects is limited, so we used four ranges with boundaries at $\pca = [0,0.15,0.5,0.85,1]$. The uncertainties were estimated from bootstrapping. The obtained values are compatible with the desired one-to-one relation, demonstrating the reliability of the computed $\pca$ and reinforcing the temperature evolution derived in Sect.~\ref{sec:fca_teff}.

The probabilities $p_{\rm Ca}$ can be therefore used as prior in the analysis of the J-PLUS photometry when models of metal-polluted atmospheres are used, properly weighting the likelihood of different spectral types and solving the degeneracy problem exposed in Sect.~\ref{sec:methods}. This will be addressed in a future work. 

\subsection{Spectroscopic confirmation of new metal-polluted white dwarfs}\label{sec:pca_gtc}
As a direct application of the derived probabilities, we selected white dwarfs with $p_{\rm Ca} \geq 0.99$ from the final sample. We obtained a total of $39$ high-confidence candidates, as summarized in Table~\ref{tab:dzcand}. We searched for information in the Montreal white dwarf database\footnote{\url{http://www.montrealwhitedwarfdatabase.org}} \citep{MWDD} and Simbad\footnote{\url{http://simbad.u-strasbg.fr/simbad}} \citep{simbad}, finding spectroscopic information for $20$ of them. All are classified as metal-polluted white dwarfs.

Thus, we have $19$ high-confidence candidates to new metal-polluted systems. We performed the observing run GTC111-21A with the Optical System for Imaging and low-Intermediate-Resolution Integrated Spectroscopy (OSIRIS) instrument at Gran Telescopio Canarias (GTC) to follow up six of these sources. The R1000B grism covering the $3\,600$ $\AA$ to $7\,500$ $\AA$ range and a long slit of $0.6^{\prime\prime}$ were used, providing a resolution of $2.1$ $\AA$ pix$^{-1}$. Three exposures per target with an on-source time ranging from $100$ s to $750$ s each, depending on the magnitude of the source, were acquired. A standard reduction, including wavelength and flux calibration, was performed. The spectra confirmed all the candidates as metal-polluted white dwarfs (Fig.~\ref{fig:dzcand}). The analysis of the abundances in these systems deserves a dedicated work.

With the addition of our new objects, $26$ of the $39$ high-confidence candidates have spectroscopic information with a purity of $100$\% for the presence of polluting metals. This was expected due to the high probability imposed for the presence of calcium absorption.

\section{Conclusions}\label{sec:conclusions}
We used the J-PLUS DR2 photometry in twelve optical bands over $2\,176$ deg$^2$ to estimate the fraction of white dwarfs with presence of \ion{Ca}{ii} H+K absorption along the cooling sequence.

The $100\ \AA$ width $J0395$ passband is sensitive to calcium absorption. We compared the photometry in the remaining eleven J-PLUS passbands using metal-free models to estimate the expected unpolluted continuum at $J0395$. This continuum is compared with the observed $J0395$ flux to estimate its equivalent width, $\ewca$. In addition, the effective temperature and the mass were estimated.

The J-PLUS + {\it Gaia} sample of $\nwd$ white dwarfs presented in \citet{clsj22pda} was analyzed. We compared our photometric measurements with the Ca/He abundances and the effective temperatures from the spectroscopic study of \citet{coutu19} for the $44$ objects in common between both samples. We found that $\ewca$ is a robust proxy for the presence of \ion{Ca}{ii} H+K absorption. In addition, the photometric temperatures are $4$\% larger than the spectroscopic ones.

Accounting for the limitations in our methodology and to avoid double degenerate systems, a total of $4\,399$ white dwarfs with effective temperature $30\,000 > T_{\rm eff} > 5\,500$ K and mass $M > 0.45$ $M_{\odot}$ were finally used. Their $\ewca$ distribution was modeled with two populations, corresponding to polluted and non-polluted systems, to estimate the fraction of calcium white dwarfs ($f_{\rm Ca}$) as a function of $T_{\rm eff}$. The observational errors were accounted for in the process adopting a Bayesian formalism.

We found that the fraction of calcium white dwarfs varies along the cooling sequence, increasing from $f_{\rm Ca} \approx 0$ at $T_{\rm eff} = 13\,500$ K to $f_{\rm Ca} \approx 0.15$ at $T_{\rm eff} = 5\,500$ K. This trend reflects the selection function of metal-polluted systems in the optical, with an increase in the \ion{Ca}{ii} H+K equivalent width at a fixed abundance towards lower effective temperature. We compared our results with the fractions obtained from the 40 pc spectroscopic sample and the SDSS DR16 spectra gathered by GF21. The J-PLUS trend is also present in the 40 pc sample, however SDSS presents a deficit of metal-polluted objects at $T_{\rm eff} > 12\,000$ K, probably due to the color selection of SDSS spectroscopic targets. 

Finally, the probability of having calcium absorption for a white dwarf, $\pca$, was computed. The comparison with the spectral classification shows that the derived probabilities are reliable, providing the right fraction of true positives for a given $\pca$. We selected $39$ white dwarfs with $\pca > 0.99$, $20$ of which have available spectra from the literature and six were followed up with OSIRIS at GTC. We confirmed all these $26$ objects as metal-polluted systems.

Our results imply that the distributions of accretion rates and metal abundances of the white dwarf population based on SDSS spectroscopy are affected by the color selection. The ongoing and future spectroscopic observations from SDSS-V Milky Way mapper, WEAVE, DESI, and 4-MOST will gather homogeneous spectroscopic data sets to overcome the current SDSS limitations. In addition, the low-resolution ($R \sim 30 - 90$) blue photometer/red photometer (BP/RP) spectra from {\it Gaia} DR3 \citep{gaiadr3_bprp, gaiadr3_calib} and future data releases will also provide valuable information for the white dwarf population \citep{carrasco14_wd, gaiadr3_synphot,jimenezesteban23,torres23,garciazamora23,vincent24}.

As a final consideration, a precise characterization for the selection function of the GF21 catalog is also essential. This catalog is based on {\it Gaia} DR3 and is used by the aforementioned spectroscopic projects as parent sample to select their targets. We envision the J-PAS photometric data as the most promising tool to derive the selection function of present and future {\it Gaia}-based catalogs for white dwarfs. As illustrated by \citet{clsj22mjpaswd} with miniJPAS data, the first square degree observed with the $54$ narrow bands of J-PAS to nominal depth \citep{minijpas}, the  J-PAS low-resolution ($R \sim 50$) photo-spectra are able to provide atmospheric compositions (H- and He-dominated) and detect metal pollution down to $r = 21.5$ mag for $22\,000 < T_{\rm eff} < 7\,000$ K white dwarfs. Moreover, a pure photometric analysis using J-PAS data alone is able to distinguish between white dwarfs and extragalactic quasars. Because J-PAS is deeper than {\it Gaia}, it will provide an independent dataset to test the {\it Gaia} selection function of metal-polluted white dwarfs and unique, faint candidates for spectroscopic follow up.


\begin{acknowledgements}
We dedicate this paper to the memory of our six IAC colleagues and friends who met with a fatal accident in Piedra de los Cochinos, Tenerife, in February 2007, with  special thanks to Maurizio Panniello, whose teachings of \texttt{python} were so important for this paper.


Based on observations made with the JAST80 telescope at the Observatorio Astrof\'{\i}sico de Javalambre (OAJ), in Teruel, owned, managed, and operated by the Centro de Estudios de F\'{\i}sica del  Cosmos de Arag\'on. We acknowledge the OAJ Data Processing and Archiving Unit (UPAD, \citealt{upad}) for reducing and calibrating the OAJ data used in this work.

Partially based on observations made with the Gran Telescopio Canarias (GTC), installed at the Spanish Observatorio del Roque de los Muchachos of the Instituto de Astrof\'{\i}sica de Canarias, on the island of La Palma, and with the instrument OSIRIS, built by a Consortium led by the Instituto de Astrof\'{\i}sica de Canarias in collaboration with the Instituto de Astronom\'{\i}a of the Universidad Aut\'onoma de M\'exico. OSIRIS was funded by GRANTECAN and the National Plan of Astronomy and Astrophysics of the Spanish Government.

Funding for the J-PLUS Project has been provided by the Governments of Spain and Arag\'on through the Fondo de Inversiones de Teruel; the Aragonese Government through the Research Groups E96, E103, E16\_17R, E16\_20R, and E16\_23R; the Spanish Ministry of Science and Innovation (MCIN/AEI/10.13039/501100011033 y FEDER, Una manera de hacer Europa) with grants PID2021-124918NB-C41, PID2021-124918NB-C42, PID2021-124918NA-C43, and PID2021-124918NB-C44; the Spanish Ministry of Science, Innovation and Universities (MCIU/AEI/FEDER, UE) with grants PGC2018-097585-B-C21 and PGC2018-097585-B-C22; the Spanish Ministry of Economy and Competitiveness (MINECO) under AYA2015-66211-C2-1-P, AYA2015-66211-C2-2, AYA2012-30789, and ICTS-2009-14; and European FEDER funding (FCDD10-4E-867, FCDD13-4E-2685). The Brazilian agencies FINEP, FAPESP, and the National Observatory of Brazil have also contributed to this project.

P.-E.~T and M.~W.~O have received funding from the European Research Council under the European Union’s Horizon 2020 research and innovation programme number 101002408 (MOS100PC).

The work by J.M. Carrasco was (partially) supported by the Spanish MICIN/AEI/10.13039/501100011033 and by "ERDF A way of making Europe" by the “European Union” through grant PID2021-122842OB-C21, and the Institute of Cosmos Sciences University of Barcelona (ICCUB, Unidad de Excelencia ’Mar\'{\i}a de Maeztu’) through grant CEX2019-000918-M.

A.~E., A.~d.~P., H.~D.~S., and J.~A.~F.~O. acknowledge the financial support from the Spanish Ministry of Science and Innovation and the European Union - NextGenerationEU through the Recovery and Resilience Facility project ICTS-MRR-2021-03-CEFCA.

F.~J.~E acknowledge financial support from MCIN/AEI/10.13039/501100011033 through grant PID2020-112949GB-I00.

A.~R.~M. acknowledges support by the Spanish MINECO grant PID2020-117252GB-I00 and by the AGAUR/Generalitat de Catalunya grant SGR-386/2021.

J.~V. acknowledges the technical members of the UPAD for their invaluable work: Juan Castillo, Javier Hern\'andez, \'Angel L\'opez, Alberto Moreno, David Muniesa, and Hector Vives.

This work has made use of data from the European Space Agency (ESA) mission
{\it Gaia} (\url{https://www.cosmos.esa.int/gaia}), processed by the {\it Gaia} Data Processing and Analysis Consortium (DPAC, \url{https://www.cosmos.esa.int/web/gaia/dpac/consortium}). Funding for the DPAC has been provided by national institutions, in particular the institutions participating in the {\it Gaia} Multilateral Agreement.

Funding for SDSS-III has been provided by the Alfred P. Sloan Foundation, the Participating Institutions, the National Science Foundation, and the U.S. Department of Energy Office of Science. The SDSS-III web site is \url{http://www.sdss3.org/}.

SDSS-III is managed by the Astrophysical Research Consortium for the Participating Institutions of the SDSS-III Collaboration including the University of Arizona, the Brazilian Participation Group, Brookhaven National Laboratory, Carnegie Mellon University, University of Florida, the French Participation Group, the German Participation Group, Harvard University, the Instituto de Astrofisica de Canarias, the Michigan State/Notre Dame/JINA Participation Group, Johns Hopkins University, Lawrence Berkeley National Laboratory, Max Planck Institute for Astrophysics, Max Planck Institute for Extraterrestrial Physics, New Mexico State University, New York University, Ohio State University, Pennsylvania State University, University of Portsmouth, Princeton University, the Spanish Participation Group, University of Tokyo, University of Utah, Vanderbilt University, University of Virginia, University of Washington, and Yale University.

Funding for the Sloan Digital Sky Survey IV has been provided by the  Alfred P. Sloan Foundation, the U.S. Department of Energy Office of Science, and the Participating  Institutions. SDSS-IV acknowledges support and resources from the Center for High Performance Computing  at the  University of Utah. The SDSS  website is \url{www.sdss.org}. SDSS-IV is managed by the Astrophysical Research Consortium for the Participating Institutions of the SDSS Collaboration including the Brazilian Participation Group, the Carnegie Institution for Science, Carnegie Mellon University, Center for Astrophysics | Harvard \& Smithsonian, the Chilean Participation Group, the French Participation Group, Instituto de Astrof\'isica de Canarias, The Johns Hopkins University, Kavli Institute for the Physics and Mathematics of the Universe (IPMU) / University of Tokyo, the Korean Participation Group, Lawrence Berkeley National Laboratory, Leibniz Institut f\"ur Astrophysik Potsdam (AIP),  Max-Planck-Institut f\"ur Astronomie (MPIA Heidelberg), Max-Planck-Institut f\"ur Astrophysik (MPA Garching), Max-Planck-Institut f\"ur Extraterrestrische Physik (MPE), National Astronomical Observatories of China, New Mexico State University, New York University, University of Notre Dame, Observat\'ario Nacional / MCTI, The Ohio State University, Pennsylvania State University, Shanghai Astronomical Observatory, United Kingdom Participation Group, Universidad Nacional Aut\'onoma de M\'exico, University of Arizona, University of Colorado Boulder, University of Oxford, University of Portsmouth, University of Utah, University of Virginia, University of Washington, University of Wisconsin, Vanderbilt University, and Yale University.

This research made use of \texttt{Astropy}, a community-developed core \texttt{Python} package for Astronomy \citep{astropy}, and \texttt{Matplotlib}, a 2D graphics package used for \texttt{Python} for publication-quality image generation across user interfaces and operating systems \citep{pylab}.
\end{acknowledgements}

\bibliographystyle{aa}
\bibliography{biblio}

\begin{thebibliography}{87}
\expandafter\ifx\csname natexlab\endcsname\relax\def\natexlab#1{#1}\fi

\bibitem[{{Ahumada} {et~al.}(2020){Ahumada}, {Prieto}, {Almeida}, {Anders},
  {Anderson}, {Andrews}, {Anguiano}, {Arcodia}, {Armengaud}, {Aubert}, {Avila},
  {Avila-Reese}, {Badenes}, {Balland}, {Barger}, {Barrera-Ballesteros}, {Basu},
  {Bautista}, {Beaton}, {Beers}, {Benavides}, {Bender}, {Bernardi}, {Bershady},
  {Beutler}, {Bidin}, {Bird}, {Bizyaev}, {Blanc}, {Blanton}, {Boquien},
  {Borissova}, {Bovy}, {Brandt}, {Brinkmann}, {Brownstein}, {Bundy}, {Bureau},
  {Burgasser}, {Burtin}, {Cano-D{\'\i}az}, {Capasso}, {Cappellari}, {Carrera},
  {Chabanier}, {Chaplin}, {Chapman}, {Cherinka}, {Chiappini}, {Doohyun Choi},
  {Chojnowski}, {Chung}, {Clerc}, {Coffey}, {Comerford}, {Comparat}, {da
  Costa}, {Cousinou}, {Covey}, {Crane}, {Cunha}, {Ilha}, {Dai}, {Damsted},
  {Darling}, {Davidson}, {Davies}, {Dawson}, {De}, {de la Macorra}, {De Lee},
  {Queiroz}, {Deconto Machado}, {de la Torre}, {Dell'Agli}, {du Mas des
  Bourboux}, {Diamond-Stanic}, {Dillon}, {Donor}, {Drory}, {Duckworth},
  {Dwelly}, {Ebelke}, {Eftekharzadeh}, {Davis Eigenbrot}, {Elsworth},
  {Eracleous}, {Erfanianfar}, {Escoffier}, {Fan}, {Farr},
  {Fern{\'a}ndez-Trincado}, {Feuillet}, {Finoguenov}, {Fofie},
  {Fraser-McKelvie}, {Frinchaboy}, {Fromenteau}, {Fu}, {Galbany}, {Garcia},
  {Garc{\'\i}a-Hern{\'a}ndez}, {Oehmichen}, {Ge}, {Maia}, {Geisler}, {Gelfand},
  {Goddy}, {Gonzalez-Perez}, {Grabowski}, {Green}, {Grier}, {Guo}, {Guy},
  {Harding}, {Hasselquist}, {Hawken}, {Hayes}, {Hearty}, {Hekker}, {Hogg},
  {Holtzman}, {Horta}, {Hou}, {Hsieh}, {Huber}, {Hunt}, {Chitham}, {Imig},
  {Jaber}, {Angel}, {Johnson}, {Jones}, {J{\"o}nsson}, {Jullo}, {Kim},
  {Kinemuchi}, {Kirkpatrick}, {Kite}, {Klaene}, {Kneib}, {Kollmeier}, {Kong},
  {Kounkel}, {Krishnarao}, {Lacerna}, {Lan}, {Lane}, {Law}, {Le Goff}, {Leung},
  {Lewis}, {Li}, {Lian}, {Lin}, {Long}, {Longa-Pe{\~n}a}, {Lundgren}, {Lyke},
  {Ted Mackereth}, {MacLeod}, {Majewski}, {Manchado}, {Maraston}, {Martini},
  {Masseron}, {Masters}, {Mathur}, {McDermid}, {Merloni}, {Merrifield},
  {M{\'e}sz{\'a}ros}, {Miglio}, {Minniti}, {Minsley}, {Miyaji}, {Mohammad},
  {Mosser}, {Mueller}, {Muna}, {Mu{\~n}oz-Guti{\'e}rrez}, {Myers}, {Nadathur},
  {Nair}, {Nandra}, {do Nascimento}, {Nevin}, {Newman}, {Nidever}, {Nitschelm},
  {Noterdaeme}, {O'Connell}, {Olmstead}, {Oravetz}, {Oravetz}, {Osorio},
  {Pace}, {Padilla}, {Palanque-Delabrouille}, {Palicio}, {Pan}, {Pan},
  {Parker}, {Paviot}, {Peirani}, {Ram{\'r}ez}, {Penny}, {Percival},
  {Perez-Fournon}, {P{\'e}rez-R{\`a}fols}, {Petitjean}, {Pieri},
  {Pinsonneault}, {Poovelil}, {Povick}, {Prakash}, {Price-Whelan}, {Raddick},
  {Raichoor}, {Ray}, {Rembold}, {Rezaie}, {Riffel}, {Riffel}, {Rix}, {Robin},
  {Roman-Lopes}, {Rom{\'a}n-Z{\'u}{\~n}iga}, {Rose}, {Ross}, {Rossi},
  {Rowlands}, {Rubin}, {Salvato}, {S{\'a}nchez}, {S{\'a}nchez-Menguiano},
  {S{\'a}nchez-Gallego}, {Sayres}, {Schaefer}, {Schiavon}, {Schimoia},
  {Schlafly}, {Schlegel}, {Schneider}, {Schultheis}, {Schwope}, {Seo},
  {Serenelli}, {Shafieloo}, {Shamsi}, {Shao}, {Shen}, {Shetrone}, {Shirley},
  {Aguirre}, {Simon}, {Skrutskie}, {Slosar}, {Smethurst}, {Sobeck}, {Sodi},
  {Souto}, {Stark}, {Stassun}, {Steinmetz}, {Stello}, {Stermer},
  {Storchi-Bergmann}, {Streblyanska}, {Stringfellow}, {Stutz}, {Su{\'a}rez},
  {Sun}, {Taghizadeh-Popp}, {Talbot}, {Tayar}, {Thakar}, {Theriault}, {Thomas},
  {Thomas}, {Tinker}, {Tojeiro}, {Toledo}, {Tremonti}, {Troup}, {Tuttle},
  {Unda-Sanzana}, {Valentini}, {Vargas-Gonz{\'a}lez}, {Vargas-Maga{\~n}a},
  {V{\'a}zquez-Mata}, {Vivek}, {Wake}, {Wang}, {Weaver}, {Weijmans}, {Wild},
  {Wilson}, {Wilson}, {Wolthuis}, {Wood-Vasey}, {Yan}, {Yang}, {Y{\`e}che},
  {Zamora}, {Zarrouk}, {Zasowski}, {Zhang}, {Zhao}, {Zhao}, {Zheng}, {Zheng},
  {Zhu}, \& {Zou}}]{sdss_dr16}
{Ahumada}, R., {Prieto}, C.~A., {Almeida}, A., {et~al.} 2020, \apjs, 249, 3

\bibitem[{{Allende Prieto} {et~al.}(2020){Allende Prieto}, {Cooper}, {Dey},
  {G{\"a}nsicke}, {Koposov}, {Li}, {Manser}, {Nidever}, {Rockosi}, {Wang},
  {Aguado}, {Blum}, {Brooks}, {Eisenstein}, {Duan}, {Eftekharzadeh},
  {Gazta{\~n}aga}, {Kehoe}, {Landriau}, {Lee}, {Levi}, {Meisner}, {Myers},
  {Najita}, {Olsen}, {Palanque-Delabrouille}, {Poppett}, {Prada}, {Schlegel},
  {Schubnell}, {Tarl{\'e}}, {Valluri}, {Wechsler}, \& {Y{\`e}che}}]{desi_mws}
{Allende Prieto}, C., {Cooper}, A.~P., {Dey}, A., {et~al.} 2020, Research Notes
  of the American Astronomical Society, 4, 188

\bibitem[{{Astropy Collaboration} {et~al.}(2013){Astropy Collaboration},
  {Robitaille}, {Tollerud}, {Greenfield}, {Droettboom}, {Bray}, {Aldcroft},
  {Davis}, {Ginsburg}, {Price-Whelan}, {Kerzendorf}, {Conley}, {Crighton},
  {Barbary}, {Muna}, {Ferguson}, {Grollier}, {Parikh}, {Nair}, {Unther},
  {Deil}, {Woillez}, {Conseil}, {Kramer}, {Turner}, {Singer}, {Fox}, {Weaver},
  {Zabalza}, {Edwards}, {Azalee Bostroem}, {Burke}, {Casey}, {Crawford},
  {Dencheva}, {Ely}, {Jenness}, {Labrie}, {Lim}, {Pierfederici}, {Pontzen},
  {Ptak}, {Refsdal}, {Servillat}, \& {Streicher}}]{astropy}
{Astropy Collaboration}, {Robitaille}, T.~P., {Tollerud}, E.~J., {et~al.} 2013,
  \aap, 558, A33

\bibitem[{{Ben{\'{\i}}tez} {et~al.}(2014){Ben{\'{\i}}tez}, {Dupke}, {Moles},
  {Sodre}, {Cenarro}, {Marin-Franch}, {Taylor}, {Cristobal}, {Fernandez-Soto},
  {Mendes de Oliveira}, {Cepa-Nogue}, {Abramo}, {Alcaniz}, {Overzier},
  {Hernandez-Monteagudo}, {Alfaro}, {Kanaan}, {Carvano}, {Reis}, {Martinez
  Gonzalez}, {Ascaso}, {Ballesteros}, {Xavier}, {Varela}, {Ederoclite},
  {Vazquez Ramio}, {Broadhurst}, {Cypriano}, {Angulo}, {Diego}, {Zandivarez},
  {Diaz}, {Melchior}, {Umetsu}, {Spinelli}, {Zitrin}, {Coe}, {Yepes}, {Vielva},
  {Sahni}, {Marcos-Caballero}, {Shu Kitaura}, {Maroto}, {Masip}, {Tsujikawa},
  {Carneiro}, {Gonzalez Nuevo}, {Carvalho}, {Reboucas}, {Carvalho}, {Abdalla},
  {Bernui}, {Pigozzo}, {Ferreira}, {Chandrachani Devi}, {Bengaly}, {Campista},
  {Amorim}, {Asari}, {Bongiovanni}, {Bonoli}, {Bruzual}, {Cardiel}, {Cava},
  {Cid Fernandes}, {Coelho}, {Cortesi}, {Delgado}, {Diaz Garcia}, {Espinosa},
  {Galliano}, {Gonzalez-Serrano}, {Falcon-Barroso}, {Fritz}, {Fernandes},
  {Gorgas}, {Hoyos}, {Jimenez-Teja}, {Lopez-Aguerri}, {Lopez-San Juan},
  {Mateus}, {Molino}, {Novais}, {OMill}, {Oteo}, {Perez-Gonzalez}, {Poggianti},
  {Proctor}, {Ricciardelli}, {Sanchez-Blazquez}, {Storchi-Bergmann}, {Telles},
  {Schoennell}, {Trujillo}, {Vazdekis}, {Viironen}, {Daflon},
  {Aparicio-Villegas}, {Rocha}, {Ribeiro}, {Borges}, {Martins}, {Marcolino},
  {Martinez-Delgado}, {Perez-Torres}, {Siffert}, {Calvao}, {Sako}, {Kessler},
  {Alvarez-Candal}, {De Pra}, {Roig}, {Lazzaro}, {Gorosabel}, {Lopes de
  Oliveira}, {Lima-Neto}, {Irwin}, {Liu}, {Alvarez}, {Balmes}, {Chueca},
  {Costa-Duarte}, {da Costa}, {Dantas}, {Diaz}, {Fabregat}, {Ferrari},
  {Gavela}, {Gracia}, {Gruel}, {Gutierrez}, {Guzman}, {Hernandez-Fernandez},
  {Herranz}, {Hurtado-Gil}, {Jablonsky}, {Laporte}, {Le Tiran}, {Licandro},
  {Lima}, {Martin}, {Martinez}, {Montero}, {Penteado}, {Pereira}, {Peris},
  {Quilis}, {Sanchez-Portal}, {Soja}, {Solano}, {Torra}, \&
  {Valdivielso}}]{jpas}
{Ben{\'{\i}}tez}, N., {Dupke}, R., {Moles}, M., {et~al.} 2014
  [\eprint[arXiv]{1403.5237}]

\bibitem[{{Bergeron} {et~al.}(2019){Bergeron}, {Dufour}, {Fontaine}, {Coutu},
  {Blouin}, {Genest-Beaulieu}, {B{\'e}dard}, \& {Rolland }}]{bergeron19}
{Bergeron}, P., {Dufour}, P., {Fontaine}, G., {et~al.} 2019, \apj, 876, 67

\bibitem[{{Bertin} \& {Arnouts}(1996)}]{sextractor}
{Bertin}, E. \& {Arnouts}, S. 1996, \aaps, 117, 393

\bibitem[{{Blouin}(2020)}]{blouin20}
{Blouin}, S. 2020, \mnras, 496, 1881

\bibitem[{{Blouin} \& {Xu}(2022)}]{blouin22}
{Blouin}, S. \& {Xu}, S. 2022, \mnras, 510, 1059

\bibitem[{{Bonoli} {et~al.}(2021){Bonoli}, {Mar{\'\i}n-Franch}, {Varela},
  {V{\'a}zquez Rami{\'o}}, {Abramo}, {Cenarro}, {Dupke}, {V{\'\i}lchez},
  {Crist{\'o}bal-Hornillos}, {Gonz{\'a}lez Delgado},
  {Hern{\'a}ndez-Monteagudo}, {L{\'o}pez-Sanjuan}, {Muniesa}, {Civera},
  {Ederoclite}, {Hern{\'a}n-Caballero}, {Marra}, {Baqui}, {Cortesi},
  {Cypriano}, {Daflon}, {de Amorim}, {D{\'\i}az-Garc{\'\i}a}, {Diego},
  {Mart{\'\i}nez-Solaeche}, {P{\'e}rez}, {Placco}, {Prada}, {Queiroz},
  {Alcaniz}, {Alvarez-Candal}, {Cepa}, {Maroto}, {Roig}, {Siffert}, {Taylor},
  {Benitez}, {Moles}, {Sodr{\'e}}, {Carneiro}, {Mendes de Oliveira}, {Abdalla},
  {Angulo}, {Aparicio Resco}, {Balaguera-Antol{\'\i}nez}, {Ballesteros},
  {Brito-Silva}, {Broadhurst}, {Carrasco}, {Castro}, {Cid Fernandes}, {Coelho},
  {de Melo}, {Doubrawa}, {Fernandez-Soto}, {Ferrari}, {Finoguenov},
  {Garc{\'\i}a-Benito}, {Iglesias-P{\'a}ramo}, {Jim{\'e}nez-Teja}, {Kitaura},
  {Laur}, {Lopes}, {Lucatelli}, {Mart{\'\i}nez}, {Maturi}, {Overzier},
  {Pigozzo}, {Quartin}, {Rodr{\'\i}guez-Mart{\'\i}n}, {Salzano}, {Tamm},
  {Tempel}, {Umetsu}, {Valdivielso}, {von Marttens}, {Zitrin},
  {D{\'\i}az-Mart{\'\i}n}, {L{\'o}pez-Alegre}, {L{\'o}pez-Sainz},
  {Yanes-D{\'\i}az}, {Rueda-Teruel}, {Rueda-Teruel}, {Abril Iba{\~n}ez}, {L
  Ant{\'o}n Bravo}, {Bello Ferrer}, {Bielsa}, {Casino}, {Castillo}, {Chueca},
  {Cuesta}, {Garzar{\'a}n Calderaro}, {Iglesias-Marzoa}, {{\'I}niguez},
  {Lamadrid Gutierrez}, {Lopez-Martinez}, {Lozano-P{\'e}rez}, {Ma{\'\i}cas
  Sacrist{\'a}n}, {Molina-Ib{\'a}{\~n}ez}, {Moreno-Signes}, {Rodr{\'\i}guez
  Llano}, {Royo Navarro}, {Tilve Rua}, {Andrade}, {Alfaro}, {Akras},
  {Arnalte-Mur}, {Ascaso}, {Barbosa}, {Beltr{\'a}n Jim{\'e}nez}, {Benetti},
  {Bengaly}, {Bernui}, {Blanco-Pillado}, {Borges Fernandes}, {Bregman},
  {Bruzual}, {Calderone}, {Carvano}, {Casarini}, {Chaves-Montero},
  {Chies-Santos}, {Coutinho de Carvalho}, {Dimauro}, {Duarte Puertas},
  {Figueruelo}, {Gonz{\'a}lez-Serrano}, {Guerrero}, {Gurung-L{\'o}pez},
  {Herranz}, {Huertas-Company}, {Irwin}, {Izquierdo-Villalba}, {Kanaan},
  {Kehrig}, {Kirkpatrick}, {Lim}, {Lopes}, {Lopes de Oliveira},
  {Marcos-Caballero}, {Mart{\'\i}nez-Delgado}, {Mart{\'\i}nez-Gonz{\'a}lez},
  {Mart{\'\i}nez-Somonte}, {Oliveira}, {Orsi}, {Penna-Lima}, {Reis}, {Spinoso},
  {Tsujikawa}, {Vielva}, {Vitorelli}, {Xia}, {Yuan}, {Arroyo-Polonio},
  {Dantas}, {Galarza}, {Gon{\c{c}}alves}, {Gon{\c{c}}alves}, {Gonzalez},
  {Gonzalez}, {Greisel}, {Jim{\'e}nez-Esteban}, {Landim}, {Lazzaro}, {Magris},
  {Monteiro-Oliveira}, {Pereira}, {Rebou{\c{c}}as}, {Rodriguez-Espinosa},
  {Santos da Costa}, \& {Telles}}]{minijpas}
{Bonoli}, S., {Mar{\'\i}n-Franch}, A., {Varela}, J., {et~al.} 2021, \aap, 653,
  A31

\bibitem[{{Cameron}(2011)}]{cameron11}
{Cameron}, E. 2011, \pasa, 28, 128

\bibitem[{{Carrasco} {et~al.}(2014){Carrasco}, {Catal{\'a}n}, {Jordi},
  {Tremblay}, {Napiwotzki}, {Luri}, {Robin}, \& {Kowalski}}]{carrasco14_wd}
{Carrasco}, J.~M., {Catal{\'a}n}, S., {Jordi}, C., {et~al.} 2014, \aap, 565,
  A11

\bibitem[{{Carter} {et~al.}(2013){Carter}, {Marsh}, {Steeghs}, {Groot},
  {Nelemans}, {Levitan}, {Rau}, {Copperwheat}, {Kupfer}, \&
  {Roelofs}}]{carter13}
{Carter}, P.~J., {Marsh}, T.~R., {Steeghs}, D., {et~al.} 2013, \mnras, 429,
  2143

\bibitem[{{Cenarro} {et~al.}(2019){Cenarro}, {Moles},
  {Crist{\'o}bal-Hornillos}, {Mar{\'\i}n-Franch}, {Ederoclite}, {Varela},
  {L{\'o}pez-Sanjuan}, {Hern{\'a}ndez-Monteagudo}, {Angulo}, {V{\'a}zquez
  Rami{\'o}}, {Viironen}, {Bonoli}, {Orsi}, {Hurier}, {San Roman}, {Greisel},
  {Vilella-Rojo}, {D{\'\i}az-Garc{\'\i}a}, {Logro{\~n}o-Garc{\'\i}a},
  {Gurung-L{\'o}pez}, {Spinoso}, {Izquierdo-Villalba}, {Aguerri}, {Allende
  Prieto}, {Bonatto}, {Carvano}, {Chies-Santos}, {Daflon}, {Dupke},
  {Falc{\'o}n-Barroso}, {Gon{\c{c}}alves}, {Jim{\'e}nez-Teja}, {Molino},
  {Placco}, {Solano}, {Whitten}, {Abril}, {Ant{\'o}n}, {Bello}, {Bielsa de
  Toledo}, {Castillo-Ram{\'\i}rez}, {Chueca}, {Civera},
  {D{\'\i}az-Mart{\'\i}n}, {Dom{\'\i}nguez-Mart{\'\i}nez},
  {Garzar{\'a}n-Calderaro}, {Hern{\'a}ndez-Fuertes}, {Iglesias-Marzoa},
  {I{\~n}iguez}, {Jim{\'e}nez Ruiz}, {Kruuse}, {Lamadrid}, {Lasso-Cabrera},
  {L{\'o}pez-Alegre}, {L{\'o}pez-Sainz}, {Ma{\'\i}cas}, {Moreno-Signes},
  {Muniesa}, {Rodr{\'\i}guez-Llano}, {Rueda-Teruel}, {Rueda-Teruel},
  {Soriano-Lagu{\'\i}a}, {Tilve}, {Valdivielso}, {Yanes-D{\'\i}az}, {Alcaniz},
  {Mendes de Oliveira}, {Sodr{\'e}}, {Coelho}, {Lopes de Oliveira}, {Tamm},
  {Xavier}, {Abramo}, {Akras}, {Alfaro}, {Alvarez-Cand al}, {Ascaso},
  {Beasley}, {Beers}, {Borges Fernandes}, {Bruzual}, {Buzzo}, {Carrasco},
  {Cepa}, {Cortesi}, {Costa-Duarte}, {De Pr{\'a}}, {Favole}, {Galarza},
  {Galbany}, {Garcia}, {Gonz{\'a}lez Delgado}, {Gonz{\'a}lez-Serrano},
  {Guti{\'e}rrez-Soto}, {Hernandez-Jimenez}, {Kanaan}, {Kuncarayakti},
  {Landim}, {Laur}, {Licandro}, {Lima Neto}, {Lyman}, {Ma{\'\i}z
  Apell{\'a}niz}, {Miralda-Escud{\'e}}, {Morate}, {Nogueira-Cavalcante},
  {Novais}, {Oncins}, {Oteo}, {Overzier}, {Pereira}, {Rebassa-Mansergas},
  {Reis}, {Roig}, {Sako}, {Salvador-Rusi{\~n}ol}, {Sampedro},
  {S{\'a}nchez-Bl{\'a}zquez}, {Santos}, {Schmidtobreick}, {Siffert}, {Telles},
  \& {Vilchez}}]{cenarro19}
{Cenarro}, A.~J., {Moles}, M., {Crist{\'o}bal-Hornillos}, D., {et~al.} 2019,
  \aap, 622, A176

\bibitem[{{Cenarro} {et~al.}(2014){Cenarro}, {Moles}, {Mar{\'{\i}}n-Franch},
  {Crist{\'o}bal-Hornillos}, {Yanes D{\'{\i}}az}, {Ederoclite}, {Varela},
  {V{\'a}zquez-Rami{\'o}}, {Valdivielso}, {Ben{\'{\i}}tez}, {Cepa}, {Dupke},
  {Fern{\'a}ndez-Soto}, {Mendes de Oliveira}, {Sodr{\'e}}, {Taylor},
  {Rueda-Teruel}, {Rueda-Teruel}, {Luis-Simoes}, {Chueca}, {Ant{\'o}n},
  {Bello}, {D{\'{\i}}az-Mart{\'{\i}}n}, {Guill{\'e}n-Civera},
  {Hern{\'a}ndez-Fuertes}, {Iglesias-Marzoa}, {Jim{\'e}nez-Mej{\'{\i}}as},
  {Lasso-Cabrera}, {L{\'o}pez-Alegre}, {L{\'o}pez-Sainz},
  {Rodr{\'{\i}}guez-Hern{\'a}ndez}, {Su{\'a}rez}, {Lamadrid}, {Ma{\'{\i}}cas},
  {Abril-Iba{\~n}ez}, {Tilve}, \& {Rodr{\'{\i}}guez-Llano}}]{oaj}
{Cenarro}, A.~J., {Moles}, M., {Mar{\'{\i}}n-Franch}, A., {et~al.} 2014, in
  \procspie, Vol. 9149, Observatory Operations: Strategies, Processes, and
  Systems V, 91491I

\bibitem[{{Chiappini} {et~al.}(2019){Chiappini}, {Minchev}, {Starkenburg},
  {Anders}, {Gentile Fusillo}, {Gerhard}, {Guiglion}, {Khalatyan},
  {Kordopatis}, {Lemasle}, {Matijevic}, {Queiroz}, {Schwope}, {Steinmetz},
  {Storm}, {Traven}, {Tremblay}, {Valentini}, {Andrae}, {Arentsen}, {Asplund},
  {Bensby}, {Bergemann}, {Casagrande}, {Church}, {Cescutti}, {Feltzing},
  {Fouesneau}, {Grebel}, {Kovalev}, {McMillan}, {Monari}, {Rybizki}, {Ryde},
  {Rix}, {Walton}, {Xiang}, {Zucker}, \& {4MIDABLE-Lr Team}}]{4most_wd}
{Chiappini}, C., {Minchev}, I., {Starkenburg}, E., {et~al.} 2019, The
  Messenger, 175, 30

\bibitem[{{Coutu} {et~al.}(2019){Coutu}, {Dufour}, {Bergeron}, {Blouin},
  {Loranger}, {Allard}, \& {Dunlap}}]{coutu19}
{Coutu}, S., {Dufour}, P., {Bergeron}, P., {et~al.} 2019, \apj, 885, 74

\bibitem[{{Crist{\'o}bal-Hornillos} {et~al.}(2012){Crist{\'o}bal-Hornillos},
  {Gruel}, {Varela}, {L{\'o}pez-Sainz}, {Ederoclite}, {Moles}, {Cenarro},
  {Mar{\'{\i}}n-Franch}, {Hern{\'a}ndez-Fuertes}, {Yanes-D{\'{\i}}az},
  {Chueca}, {Rueda-Teruel}, {Rueda-Teruel}, \& {Luis-Simoes}}]{upad}
{Crist{\'o}bal-Hornillos}, D., {Gruel}, N., {Varela}, J., {et~al.} 2012, in
  SPIE CS, Vol. 8451

\bibitem[{{Cukanovaite} {et~al.}(2018){Cukanovaite}, {Tremblay}, {Freytag},
  {Ludwig}, \& {Bergeron}}]{cukanovaite18}
{Cukanovaite}, E., {Tremblay}, P.~E., {Freytag}, B., {Ludwig}, H.~G., \&
  {Bergeron}, P. 2018, \mnras, 481, 1522

\bibitem[{{Cukanovaite} {et~al.}(2019){Cukanovaite}, {Tremblay}, {Freytag},
  {Ludwig}, {Fontaine}, {Brassard}, {Toloza}, \& {Koester}}]{cukanovaite19}
{Cukanovaite}, E., {Tremblay}, P.~E., {Freytag}, B., {et~al.} 2019, \mnras,
  490, 1010

\bibitem[{{Dalton} {et~al.}(2012){Dalton}, {Trager}, {Abrams}, {Carter},
  {Bonifacio}, {Aguerri}, {MacIntosh}, {Evans}, {Lewis}, {Navarro}, {Agocs},
  {Dee}, {Rousset}, {Tosh}, {Middleton}, {Pragt}, {Terrett}, {Brock}, {Benn},
  {Verheijen}, {Cano Infantes}, {Bevil}, {Steele}, {Mottram}, {Bates},
  {Gribbin}, {Rey}, {Rodriguez}, {Delgado}, {Guinouard}, {Walton}, {Irwin},
  {Jagourel}, {Stuik}, {Gerlofsma}, {Roelfsma}, {Skillen}, {Ridings},
  {Balcells}, {Daban}, {Gouvret}, {Venema}, \& {Girard}}]{weave}
{Dalton}, G., {Trager}, S.~C., {Abrams}, D.~C., {et~al.} 2012, in Society of
  Photo-Optical Instrumentation Engineers (SPIE) Conference Series, Vol. 8446,
  Ground-based and Airborne Instrumentation for Astronomy IV, ed. I.~S.
  {McLean}, S.~K. {Ramsay}, \& H.~{Takami}, 84460P

\bibitem[{{De Angeli} {et~al.}(2023){De Angeli}, {Weiler}, {Montegriffo},
  {Evans}, {Riello}, {Andrae}, {Carrasco}, {Busso}, {Burgess}, {Cacciari},
  {Davidson}, {Harrison}, {Hodgkin}, {Jordi}, {Osborne}, {Pancino},
  {Altavilla}, {Barstow}, {Bailer-Jones}, {Bellazzini}, {Brown}, {Castellani},
  {Cowell}, {Delchambre}, {De Luise}, {Diener}, {Fabricius}, {Fouesneau},
  {Fr{\'e}mat}, {Gilmore}, {Giuffrida}, {Hambly}, {Hidalgo}, {Holland},
  {Kostrzewa-Rutkowska}, {van Leeuwen}, {Lobel}, {Marinoni}, {Miller},
  {Pagani}, {Palaversa}, {Piersimoni}, {Pulone}, {Ragaini}, {Rainer},
  {Richards}, {Rixon}, {Ruz-Mieres}, {Sanna}, {Sarro}, {Rowell}, {Sordo},
  {Walton}, \& {Yoldas}}]{gaiadr3_bprp}
{De Angeli}, F., {Weiler}, M., {Montegriffo}, P., {et~al.} 2023, \aap, 674, A2

\bibitem[{{Dufour} {et~al.}(2007){Dufour}, {Bergeron}, {Liebert}, {Harris},
  {Knapp}, {Anderson}, {Hall}, {Strauss}, {Collinge}, \& {Edwards}}]{dufour07}
{Dufour}, P., {Bergeron}, P., {Liebert}, J., {et~al.} 2007, \apj, 663, 1291

\bibitem[{{Dufour} {et~al.}(2017){Dufour}, {Blouin}, {Coutu},
  {Fortin-Archambault}, {Thibeault}, {Bergeron}, \& {Fontaine}}]{MWDD}
{Dufour}, P., {Blouin}, S., {Coutu}, S., {et~al.} 2017, in Astronomical Society
  of the Pacific Conference Series, Vol. 509, 20th European White Dwarf
  Workshop, ed. P.~E. {Tremblay}, B.~{Gaensicke}, \& T.~{Marsh}, 3

\bibitem[{{Eisenstein} {et~al.}(2006){Eisenstein}, {Liebert}, {Harris},
  {Kleinman}, {Nitta}, {Silvestri}, {Anderson}, {Barentine}, {Brewington},
  {Brinkmann}, {Harvanek}, {Krzesi{\'n}ski}, {Neilsen}, {Long}, {Schneider}, \&
  {Snedden}}]{eisenstein06_dr4}
{Eisenstein}, D.~J., {Liebert}, J., {Harris}, H.~C., {et~al.} 2006, \apjs, 167,
  40

\bibitem[{{Farihi} {et~al.}(2010){Farihi}, {Jura}, {Lee}, \&
  {Zuckerman}}]{farihi10}
{Farihi}, J., {Jura}, M., {Lee}, J.~E., \& {Zuckerman}, B. 2010, \apj, 714,
  1386

\bibitem[{{Fontaine} {et~al.}(2001){Fontaine}, {Brassard}, \&
  {Bergeron}}]{fontaine01}
{Fontaine}, G., {Brassard}, P., \& {Bergeron}, P. 2001, \pasp, 113, 409

\bibitem[{{Foreman-Mackey} {et~al.}(2013){Foreman-Mackey}, {Hogg}, {Lang}, \&
  {Goodman}}]{emcee}
{Foreman-Mackey}, D., {Hogg}, D.~W., {Lang}, D., \& {Goodman}, J. 2013, \pasp,
  125, 306

\bibitem[{{Gaia Collaboration} {et~al.}(2021){Gaia Collaboration}, {Brown},
  {Vallenari}, {Prusti}, {de Bruijne}, {Babusiaux}, {Biermann}, {Creevey},
  {Evans}, {Eyer}, {Hutton}, {Jansen}, {Jordi}, {Klioner}, {Lammers},
  {Lindegren}, {Luri}, {Mignard}, {Panem}, {Pourbaix}, {Randich}, {Sartoretti},
  {Soubiran}, {Walton}, {Arenou}, {Bailer-Jones}, {Bastian}, {Cropper},
  {Drimmel}, {Katz}, {Lattanzi}, {van Leeuwen}, {Bakker}, {Cacciari},
  {Casta{\~n}eda}, {De Angeli}, {Ducourant}, {Fabricius}, {Fouesneau},
  {Fr{\'e}mat}, {Guerra}, {Guerrier}, {Guiraud}, {Jean-Antoine Piccolo},
  {Masana}, {Messineo}, {Mowlavi}, {Nicolas}, {Nienartowicz}, {Pailler},
  {Panuzzo}, {Riclet}, {Roux}, {Seabroke}, {Sordo}, {Tanga}, {Th{\'e}venin},
  {Gracia-Abril}, {Portell}, {Teyssier}, {Altmann}, {Andrae}, {Bellas-Velidis},
  {Benson}, {Berthier}, {Blomme}, {Brugaletta}, {Burgess}, {Busso}, {Carry},
  {Cellino}, {Cheek}, {Clementini}, {Damerdji}, {Davidson}, {Delchambre},
  {Dell'Oro}, {Fern{\'a}ndez-Hern{\'a}ndez}, {Galluccio}, {Garc{\'\i}a-Lario},
  {Garcia-Reinaldos}, {Gonz{\'a}lez-N{\'u}{\~n}ez}, {Gosset}, {Haigron},
  {Halbwachs}, {Hambly}, {Harrison}, {Hatzidimitriou}, {Heiter},
  {Hern{\'a}ndez}, {Hestroffer}, {Hodgkin}, {Holl}, {Jan{\ss}en}, {Jevardat de
  Fombelle}, {Jordan}, {Krone-Martins}, {Lanzafame}, {L{\"o}ffler}, {Lorca},
  {Manteiga}, {Marchal}, {Marrese}, {Moitinho}, {Mora}, {Muinonen}, {Osborne},
  {Pancino}, {Pauwels}, {Petit}, {Recio-Blanco}, {Richards}, {Riello},
  {Rimoldini}, {Robin}, {Roegiers}, {Rybizki}, {Sarro}, {Siopis}, {Smith},
  {Sozzetti}, {Ulla}, {Utrilla}, {van Leeuwen}, {van Reeven}, {Abbas}, {Abreu
  Aramburu}, {Accart}, {Aerts}, {Aguado}, {Ajaj}, {Altavilla}, {{\'A}lvarez},
  {{\'A}lvarez Cid-Fuentes}, {Alves}, {Anderson}, {Anglada Varela}, {Antoja},
  {Audard}, {Baines}, {Baker}, {Balaguer-N{\'u}{\~n}ez}, {Balbinot}, {Balog},
  {Barache}, {Barbato}, {Barros}, {Barstow}, {Bartolom{\'e}}, {Bassilana},
  {Bauchet}, {Baudesson-Stella}, {Becciani}, {Bellazzini}, {Bernet}, {Bertone},
  {Bianchi}, {Blanco-Cuaresma}, {Boch}, {Bombrun}, {Bossini}, {Bouquillon},
  {Bragaglia}, {Bramante}, {Breedt}, {Bressan}, {Brouillet}, {Bucciarelli},
  {Burlacu}, {Busonero}, {Butkevich}, {Buzzi}, {Caffau}, {Cancelliere},
  {C{\'a}novas}, {Cantat-Gaudin}, {Carballo}, {Carlucci}, {Carnerero},
  {Carrasco}, {Casamiquela}, {Castellani}, {Castro-Ginard}, {Castro Sampol},
  {Chaoul}, {Charlot}, {Chemin}, {Chiavassa}, {Cioni}, {Comoretto}, {Cooper},
  {Cornez}, {Cowell}, {Crifo}, {Crosta}, {Crowley}, {Dafonte}, {Dapergolas},
  {David}, {David}, {de Laverny}, {De Luise}, {De March}, {De Ridder}, {de
  Souza}, {de Teodoro}, {de Torres}, {del Peloso}, {del Pozo}, {Delbo},
  {Delgado}, {Delgado}, {Delisle}, {Di Matteo}, {Diakite}, {Diener},
  {Distefano}, {Dolding}, {Eappachen}, {Edvardsson}, {Enke}, {Esquej}, {Fabre},
  {Fabrizio}, {Faigler}, {Fedorets}, {Fernique}, {Fienga}, {Figueras},
  {Fouron}, {Fragkoudi}, {Fraile}, {Franke}, {Gai}, {Garabato},
  {Garcia-Gutierrez}, {Garc{\'\i}a-Torres}, {Garofalo}, {Gavras}, {Gerlach},
  {Geyer}, {Giacobbe}, {Gilmore}, {Girona}, {Giuffrida}, {Gomel}, {Gomez},
  {Gonzalez-Santamaria}, {Gonz{\'a}lez-Vidal}, {Granvik},
  {Guti{\'e}rrez-S{\'a}nchez}, {Guy}, {Hauser}, {Haywood}, {Helmi}, {Hidalgo},
  {Hilger}, {H{\l}adczuk}, {Hobbs}, {Holland}, {Huckle}, {Jasniewicz},
  {Jonker}, {Juaristi Campillo}, {Julbe}, {Karbevska}, {Kervella}, {Khanna},
  {Kochoska}, {Kontizas}, {Kordopatis}, {Korn}, {Kostrzewa-Rutkowska},
  {Kruszy{\'n}ska}, {Lambert}, {Lanza}, {Lasne}, {Le Campion}, {Le Fustec},
  {Lebreton}, {Lebzelter}, {Leccia}, {Leclerc}, {Lecoeur-Taibi}, {Liao},
  {Licata}, {Lindstr{\o}m}, {Lister}, {Livanou}, {Lobel}, {Madrero Pardo},
  {Managau}, {Mann}, {Marchant}, {Marconi}, {Marcos Santos}, {Marinoni},
  {Marocco}, {Marshall}, {Martin Polo}, {Mart{\'\i}n-Fleitas}, {Masip},
  {Massari}, {Mastrobuono-Battisti}, {Mazeh}, {McMillan}, {Messina},
  {Michalik}, {Millar}, {Mints}, {Molina}, {Molinaro}, {Moln{\'a}r},
  {Montegriffo}, {Mor}, {Morbidelli}, {Morel}, {Morris}, {Mulone}, {Munoz},
  {Muraveva}, {Murphy}, {Musella}, {Noval}, {Ord{\'e}novic}, {Orr{\`u}},
  {Osinde}, {Pagani}, {Pagano}, {Palaversa}, {Palicio}, {Panahi}, {Pawlak},
  {Pe{\~n}alosa Esteller}, {Penttil{\"a}}, {Piersimoni}, {Pineau}, {Plachy},
  {Plum}, {Poggio}, {Poretti}, {Poujoulet}, {Pr{\v{s}}a}, {Pulone}, {Racero},
  {Ragaini}, {Rainer}, {Raiteri}, {Rambaux}, {Ramos}, {Ramos-Lerate}, {Re
  Fiorentin}, {Regibo}, {Reyl{\'e}}, {Ripepi}, {Riva}, {Rixon}, {Robichon},
  {Robin}, {Roelens}, {Rohrbasser}, {Romero-G{\'o}mez}, {Rowell}, {Royer},
  {Rybicki}, {Sadowski}, {Sagrist{\`a} Sell{\'e}s}, {Sahlmann}, {Salgado},
  {Salguero}, {Samaras}, {Sanchez Gimenez}, {Sanna}, {Santove{\~n}a},
  {Sarasso}, {Schultheis}, {Sciacca}, {Segol}, {Segovia}, {S{\'e}gransan},
  {Semeux}, {Shahaf}, {Siddiqui}, {Siebert}, {Siltala}, {Slezak}, {Smart},
  {Solano}, {Solitro}, {Souami}, {Souchay}, {Spagna}, {Spoto}, {Steele},
  {Steidelm{\"u}ller}, {Stephenson}, {S{\"u}veges}, {Szabados}, {Szegedi-Elek},
  {Taris}, {Tauran}, {Taylor}, {Teixeira}, {Thuillot}, {Tonello}, {Torra},
  {Torra}, {Turon}, {Unger}, {Vaillant}, {van Dillen}, {Vanel}, {Vecchiato},
  {Viala}, {Vicente}, {Voutsinas}, {Weiler}, {Wevers}, {Wyrzykowski}, {Yoldas},
  {Yvard}, {Zhao}, {Zorec}, {Zucker}, {Zurbach}, \& {Zwitter}}]{gaiaedr3}
{Gaia Collaboration}, {Brown}, A.~G.~A., {Vallenari}, A., {et~al.} 2021, \aap,
  649, A1

\bibitem[{{Gaia Collaboration} {et~al.}(2023){Gaia Collaboration},
  {Montegriffo}, {Bellazzini}, {De Angeli}, {Andrae}, {Barstow}, {Bossini},
  {Bragaglia}, {Burgess}, {Cacciari}, {Carrasco}, {Chornay}, {Delchambre},
  {Evans}, {Fouesneau}, {Fr{\'e}mat}, {Garabato}, {Jordi}, {Manteiga},
  {Massari}, {Palaversa}, {Pancino}, {Riello}, {Ruz Mieres}, {Sanna},
  {Santove{\~n}a}, {Sordo}, {Vallenari}, {Walton}, {Brown}, {Prusti}, {de
  Bruijne}, {Arenou}, {Babusiaux}, {Biermann}, {Creevey}, {Ducourant}, {Eyer},
  {Guerra}, {Hutton}, {Klioner}, {Lammers}, {Lindegren}, {Luri}, {Mignard},
  {Panem}, {Pourbaix}, {Randich}, {Sartoretti}, {Soubiran}, {Tanga},
  {Bailer-Jones}, {Bastian}, {Drimmel}, {Jansen}, {Katz}, {Lattanzi}, {van
  Leeuwen}, {Bakker}, {Casta{\~n}eda}, {Fabricius}, {Galluccio}, {Guerrier},
  {Heiter}, {Masana}, {Messineo}, {Mowlavi}, {Nicolas}, {Nienartowicz},
  {Pailler}, {Panuzzo}, {Riclet}, {Roux}, {Seabroke}, {Th{\'e}venin},
  {Gracia-Abril}, {Portell}, {Teyssier}, {Altmann}, {Audard}, {Bellas-Velidis},
  {Benson}, {Berthier}, {Blomme}, {Busonero}, {Busso}, {C{\'a}novas}, {Carry},
  {Cellino}, {Cheek}, {Clementini}, {Damerdji}, {Davidson}, {de Teodoro},
  {Nu{\~n}ez Campos}, {Dell'Oro}, {Esquej}, {Fern{\'a}ndez-Hern{\'a}ndez},
  {Fraile}, {Garc{\'\i}a-Lario}, {Gosset}, {Haigron}, {Halbwachs}, {Hambly},
  {Harrison}, {Hern{\'a}ndez}, {Hestroffer}, {Hodgkin}, {Holl}, {Jan{\ss}en},
  {Jevardat de Fombelle}, {Jordan}, {Krone-Martins}, {Lanzafame},
  {L{\"o}ffler}, {Marchal}, {Marrese}, {Moitinho}, {Muinonen}, {Osborne},
  {Pauwels}, {Recio-Blanco}, {Reyl{\'e}}, {Rimoldini}, {Roegiers}, {Rybizki},
  {Sarro}, {Siopis}, {Smith}, {Sozzetti}, {Utrilla}, {van Leeuwen}, {Abbas},
  {{\'A}brah{\'a}m}, {Abreu Aramburu}, {Aerts}, {Aguado}, {Ajaj},
  {Aldea-Montero}, {Altavilla}, {{\'A}lvarez}, {Alves}, {Anderson}, {Anglada
  Varela}, {Antoja}, {Baines}, {Baker}, {Balaguer-N{\'u}{\~n}ez}, {Balbinot},
  {Balog}, {Barache}, {Barbato}, {Barros}, {Bartolom{\'e}}, {Bassilana},
  {Bauchet}, {Becciani}, {Berihuete}, {Bernet}, {Bertone}, {Bianchi},
  {Binnenfeld}, {Blanco-Cuaresma}, {Boch}, {Bombrun}, {Bouquillon}, {Bramante},
  {Breedt}, {Bressan}, {Brouillet}, {Brugaletta}, {Bucciarelli}, {Burlacu},
  {Butkevich}, {Buzzi}, {Caffau}, {Cancelliere}, {Cantat-Gaudin}, {Carballo},
  {Carlucci}, {Carnerero}, {Casamiquela}, {Castellani}, {Castro-Ginard},
  {Chaoul}, {Charlot}, {Chemin}, {Chiaramida}, {Chiavassa}, {Comoretto},
  {Contursi}, {Cooper}, {Cornez}, {Cowell}, {Crifo}, {Cropper}, {Crosta},
  {Crowley}, {Dafonte}, {Dapergolas}, {David}, {de Laverny}, {De Luise}, {De
  March}, {De Ridder}, {de Souza}, {de Torres}, {del Peloso}, {del Pozo},
  {Delbo}, {Delgado}, {Delisle}, {Demouchy}, {Dharmawardena}, {Diakite},
  {Diener}, {Distefano}, {Dolding}, {Enke}, {Fabre}, {Fabrizio}, {Faigler},
  {Fedorets}, {Fernique}, {Figueras}, {Fournier}, {Fouron}, {Fragkoudi}, {Gai},
  {Garcia-Gutierrez}, {Garcia-Reinaldos}, {Garc{\'\i}a-Torres}, {Garofalo},
  {Gavel}, {Gavras}, {Gerlach}, {Geyer}, {Giacobbe}, {Gilmore}, {Girona},
  {Giuffrida}, {Gomel}, {Gomez}, {Gonz{\'a}lez-N{\'u}{\~n}ez},
  {Gonz{\'a}lez-Santamar{\'\i}a}, {Gonz{\'a}lez-Vidal}, {Granvik}, {Guillout},
  {Guiraud}, {Guti{\'e}rrez-S{\'a}nchez}, {Guy}, {Hatzidimitriou}, {Hauser},
  {Haywood}, {Helmer}, {Helmi}, {Sarmiento}, {Hidalgo}, {H{\l}adczuk}, {Hobbs},
  {Holland}, {Huckle}, {Jardine}, {Jasniewicz}, {Jean-Antoine Piccolo},
  {Jim{\'e}nez-Arranz}, {Juaristi Campillo}, {Julbe}, {Karbevska}, {Kervella},
  {Khanna}, {Kordopatis}, {Korn}, {K{\'o}sp{\'a}l}, {Kostrzewa-Rutkowska},
  {Kruszy{\'n}ska}, {Kun}, {Laizeau}, {Lambert}, {Lanza}, {Lasne}, {Le
  Campion}, {Lebreton}, {Lebzelter}, {Leccia}, {Leclerc}, {Lecoeur-Taibi},
  {Liao}, {Licata}, {Lindstr{\'o}m}, {Lister}, {Livanou}, {Lobel}, {Lorca},
  {Loup}, {Madrero Pardo}, {Magdaleno Romeo}, {Managau}, {Mann}, {Marchant},
  {Marconi}, {Marcos}, {Marcos Santos}, {Mar{\'\i}n Pina}, {Marinoni},
  {Marocco}, {Marshall}, {Martin Polo}, {Mart{\'\i}n-Fleitas}, {Marton},
  {Mary}, {Masip}, {Mastrobuono-Battisti}, {Mazeh}, {McMillan}, {Messina},
  {Michalik}, {Millar}, {Mints}, {Molina}, {Molinaro}, {Moln{\'a}r}, {Monari},
  {Mongui{\'o}}, {Montero}, {Mor}, {Mora}, {Morbidelli}, {Morel}, {Morris},
  {Muraveva}, {Murphy}, {Musella}, {Nagy}, {Noval}, {Oca{\~n}a}, {Ogden},
  {Ordenovic}, {Osinde}, {Pagani}, {Pagano}, {Palicio}, {Pallas-Quintela},
  {Panahi}, {Payne-Wardenaar}, {Pe{\~n}alosa Esteller}, {Penttil{\"a}},
  {Pichon}, {Piersimoni}, {Pineau}, {Plachy}, {Plum}, {Poggio}, {Pr{\v{s}}a},
  {Pulone}, {Racero}, {Ragaini}, {Rainer}, {Raiteri}, {Ramos}, {Ramos-Lerate},
  {Re Fiorentin}, {Regibo}, {Richards}, {Rios Diaz}, {Ripepi}, {Riva}, {Rix},
  {Rixon}, {Robichon}, {Robin}, {Robin}, {Roelens}, {Rogues}, {Rohrbasser},
  {Romero-G{\'o}mez}, {Rowell}, {Royer}, {Rybicki}, {Sadowski}, {S{\'a}ez
  N{\'u}{\~n}ez}, {Sagrist{\`a} Sell{\'e}s}, {Sahlmann}, {Salguero}, {Samaras},
  {Sanchez Gimenez}, {Sarasso}, {Schultheis}, {Sciacca}, {Segol}, {Segovia},
  {S{\'e}gransan}, {Semeux}, {Shahaf}, {Siddiqui}, {Siebert}, {Siltala},
  {Silvelo}, {Slezak}, {Slezak}, {Smart}, {Snaith}, {Solano}, {Solitro},
  {Souami}, {Souchay}, {Spagna}, {Spina}, {Spoto}, {Steele},
  {Steidelm{\"u}ller}, {Stephenson}, {S{\"u}veges}, {Surdej}, {Szabados},
  {Szegedi-Elek}, {Taris}, {Taylor}, {Teixeira}, {Tolomei}, {Tonello}, {Torra},
  {Torra}, {Torralba Elipe}, {Trabucchi}, {Tsounis}, {Turon}, {Ulla}, {Unger},
  {Vaillant}, {van Dillen}, {van Reeven}, {Vanel}, {Vecchiato}, {Viala},
  {Vicente}, {Voutsinas}, {Wevers}, {Wyrzykowski}, {Yoldas}, {Yvard}, {Zhao},
  {Zorec}, {Zucker}, \& {Zwitter}}]{gaiadr3_synphot}
{Gaia Collaboration}, {Montegriffo}, P., {Bellazzini}, M., {et~al.} 2023, \aap,
  674, A33

\bibitem[{{Gaia Collaboration} {et~al.}(2016){Gaia Collaboration}, {Prusti},
  {de Bruijne}, {Brown}, {Vallenari}, {Babusiaux}, {Bailer-Jones}, {Bastian},
  {Biermann}, {Evans}, \& et~al.}]{gaia}
{Gaia Collaboration}, {Prusti}, T., {de Bruijne}, J.~H.~J., {et~al.} 2016,
  \aap, 595, A1

\bibitem[{{Garc{\'\i}a-Zamora} {et~al.}(2023){Garc{\'\i}a-Zamora}, {Torres}, \&
  {Rebassa-Mansergas}}]{garciazamora23}
{Garc{\'\i}a-Zamora}, E.~M., {Torres}, S., \& {Rebassa-Mansergas}, A. 2023,
  \aap, 679, A127

\bibitem[{{Gentile Fusillo} {et~al.}(2015){Gentile Fusillo}, {G{\"a}nsicke}, \&
  {Greiss}}]{GF15}
{Gentile Fusillo}, N.~P., {G{\"a}nsicke}, B.~T., \& {Greiss}, S. 2015, \mnras,
  448, 2260

\bibitem[{{Gentile Fusillo} {et~al.}(2020){Gentile Fusillo}, {Tremblay},
  {Bohlin}, {Deustua}, \& {Kalirai}}]{GF20}
{Gentile Fusillo}, N.~P., {Tremblay}, P.-E., {Bohlin}, R.~C., {Deustua}, S.~E.,
  \& {Kalirai}, J.~S. 2020, \mnras, 491, 3613

\bibitem[{{Gentile Fusillo} {et~al.}(2021){Gentile Fusillo}, {Tremblay},
  {Cukanovaite}, {Vorontseva}, {Lallement}, {Hollands}, {G{\"a}nsicke},
  {Burdge}, {McCleery}, \& {Jordan}}]{GF21}
{Gentile Fusillo}, N.~P., {Tremblay}, P.~E., {Cukanovaite}, E., {et~al.} 2021,
  \mnras, 508, 3877

\bibitem[{{Giammichele} {et~al.}(2012){Giammichele}, {Bergeron}, \&
  {Dufour}}]{giammichele12}
{Giammichele}, N., {Bergeron}, P., \& {Dufour}, P. 2012, \apjs, 199, 29

\bibitem[{{Goodman} \& {Weare}(2010)}]{goodman10}
{Goodman}, J. \& {Weare}, J. 2010, Comm. App. Math. Comp. Sci., 5, 65

\bibitem[{{Hollands} {et~al.}(2018{\natexlab{a}}){Hollands}, {G{\"a}nsicke}, \&
  {Koester}}]{hollands18dz}
{Hollands}, M.~A., {G{\"a}nsicke}, B.~T., \& {Koester}, D. 2018{\natexlab{a}},
  \mnras, 477, 93

\bibitem[{{Hollands} {et~al.}(2017){Hollands}, {Koester}, {Alekseev},
  {Herbert}, \& {G{\"a}nsicke}}]{hollands17}
{Hollands}, M.~A., {Koester}, D., {Alekseev}, V., {Herbert}, E.~L., \&
  {G{\"a}nsicke}, B.~T. 2017, \mnras, 467, 4970

\bibitem[{{Hollands} {et~al.}(2018{\natexlab{b}}){Hollands}, {Tremblay},
  {G{\"a}nsicke}, {Gentile-Fusillo}, \& {Toonen}}]{hollands18}
{Hollands}, M.~A., {Tremblay}, P.~E., {G{\"a}nsicke}, B.~T., {Gentile-Fusillo},
  N.~P., \& {Toonen}, S. 2018{\natexlab{b}}, \mnras, 480, 3942

\bibitem[{Hunter(2007)}]{pylab}
Hunter, J.~D. 2007, Computing In Science \& Engineering, 9, 90

\bibitem[{{Jim{\'e}nez-Esteban} {et~al.}(2023){Jim{\'e}nez-Esteban}, {Torres},
  {Rebassa-Mansergas}, {Cruz}, {Murillo-Ojeda}, {Solano}, {Rodrigo}, \&
  {Camisassa}}]{jimenezesteban23}
{Jim{\'e}nez-Esteban}, F.~M., {Torres}, S., {Rebassa-Mansergas}, A., {et~al.}
  2023, \mnras, 518, 5106

\bibitem[{{Jura}(2003)}]{jura03}
{Jura}, M. 2003, \apjl, 584, L91

\bibitem[{{Jura}(2008)}]{jura08}
{Jura}, M. 2008, \aj, 135, 1785

\bibitem[{{Jura} \& {Young}(2014)}]{jura14}
{Jura}, M. \& {Young}, E.~D. 2014, Annual Review of Earth and Planetary
  Sciences, 42, 45

\bibitem[{{Kepler} {et~al.}(2015){Kepler}, {Pelisoli}, {Koester}, {Ourique},
  {Kleinman}, {Romero}, {Nitta}, {Eisenstein}, {Costa}, {K{\"u}lebi}, {Jordan},
  {Dufour}, {Giommi}, \& {Rebassa-Mansergas}}]{kepler15}
{Kepler}, S.~O., {Pelisoli}, I., {Koester}, D., {et~al.} 2015, \mnras, 446,
  4078

\bibitem[{{Kepler} {et~al.}(2016){Kepler}, {Pelisoli}, {Koester}, {Ourique},
  {Romero}, {Reindl}, {Kleinman}, {Eisenstein}, {Valois}, \&
  {Amaral}}]{kepler16}
{Kepler}, S.~O., {Pelisoli}, I., {Koester}, D., {et~al.} 2016, \mnras, 455,
  3413

\bibitem[{{Kepler} {et~al.}(2019){Kepler}, {Pelisoli}, {Koester}, {Reindl},
  {Geier}, {Romero}, {Ourique}, {Oliveira}, \& {Amaral}}]{kepler19}
{Kepler}, S.~O., {Pelisoli}, I., {Koester}, D., {et~al.} 2019, \mnras, 486,
  2169

\bibitem[{{Kilic} {et~al.}(2020){Kilic}, {Bergeron}, {Kosakowski}, {Brown},
  {Ag{\"u}eros}, \& {Blouin}}]{kilic20}
{Kilic}, M., {Bergeron}, P., {Kosakowski}, A., {et~al.} 2020, \apj, 898, 84

\bibitem[{{Kleinman} {et~al.}(2004){Kleinman}, {Harris}, {Eisenstein},
  {Liebert}, {Nitta}, {Krzesi{\'n}ski}, {Munn}, {Dahn}, {Hawley}, {Pier},
  {Schmidt}, {Silvestri}, {Smith}, {Szkody}, {Strauss}, {Knapp}, {Collinge},
  {Mukadam}, {Koester}, {Uomoto}, {Schlegel}, {Anderson}, {Brinkmann}, {Lamb},
  {Schneider}, \& {York}}]{kleinman04}
{Kleinman}, S.~J., {Harris}, H.~C., {Eisenstein}, D.~J., {et~al.} 2004, \apj,
  607, 426

\bibitem[{{Kleinman} {et~al.}(2013){Kleinman}, {Kepler}, {Koester}, {Pelisoli},
  {Pe{\c{c}}anha}, {Nitta}, {Costa}, {Krzesinski}, {Dufour}, {Lachapelle},
  {Bergeron}, {Yip}, {Harris}, {Eisenstein}, {Althaus}, \&
  {C{\'o}rsico}}]{kleinman13}
{Kleinman}, S.~J., {Kepler}, S.~O., {Koester}, D., {et~al.} 2013, \apjs, 204, 5

\bibitem[{{Koester}(2010)}]{koester10}
{Koester}, D. 2010, \memsai, 81, 921

\bibitem[{{Koester} {et~al.}(2014){Koester}, {G{\"a}nsicke}, \&
  {Farihi}}]{koester14}
{Koester}, D., {G{\"a}nsicke}, B.~T., \& {Farihi}, J. 2014, \aap, 566, A34

\bibitem[{{Koester} {et~al.}(2011){Koester}, {Girven}, {G{\"a}nsicke}, \&
  {Dufour}}]{koester11}
{Koester}, D., {Girven}, J., {G{\"a}nsicke}, B.~T., \& {Dufour}, P. 2011, \aap,
  530, A114

\bibitem[{{Koester} {et~al.}(2005){Koester}, {Rollenhagen}, {Napiwotzki},
  {Voss}, {Christlieb}, {Homeier}, \& {Reimers}}]{koester05}
{Koester}, D., {Rollenhagen}, K., {Napiwotzki}, R., {et~al.} 2005, \aap, 432,
  1025

\bibitem[{{Koester} \& {Wilken}(2006)}]{koester06}
{Koester}, D. \& {Wilken}, D. 2006, \aap, 453, 1051

\bibitem[{{Kollmeier} {et~al.}(2017){Kollmeier}, {Zasowski}, {Rix}, {Johns},
  {Anderson}, {Drory}, {Johnson}, {Pogge}, {Bird}, {Blanc}, {Brownstein},
  {Crane}, {De Lee}, {Klaene}, {Kreckel}, {MacDonald}, {Merloni}, {Ness},
  {O'Brien}, {Sanchez-Gallego}, {Sayres}, {Shen}, {Thakar}, {Tkachenko},
  {Aerts}, {Blanton}, {Eisenstein}, {Holtzman}, {Maoz}, {Nandra}, {Rockosi},
  {Weinberg}, {Bovy}, {Casey}, {Chaname}, {Clerc}, {Conroy}, {Eracleous},
  {G{\"a}nsicke}, {Hekker}, {Horne}, {Kauffmann}, {McQuinn}, {Pellegrini},
  {Schinnerer}, {Schlafly}, {Schwope}, {Seibert}, {Teske}, \& {van
  Saders}}]{sdssv}
{Kollmeier}, J.~A., {Zasowski}, G., {Rix}, H.-W., {et~al.} 2017,
  [ArXiv:1711.03234]

\bibitem[{{Kong} {et~al.}(2019){Kong}, {Luo}, \& {Li}}]{kong19}
{Kong}, X., {Luo}, A.~L., \& {Li}, X.-R. 2019, Research in Astronomy and
  Astrophysics, 19, 088

\bibitem[{{Limoges} {et~al.}(2015){Limoges}, {Bergeron}, \&
  {L{\'e}pine}}]{limoges15}
{Limoges}, M.~M., {Bergeron}, P., \& {L{\'e}pine}, S. 2015, \apjs, 219, 19

\bibitem[{{Limoges} {et~al.}(2013){Limoges}, {L{\'e}pine}, \&
  {Bergeron}}]{limoges13}
{Limoges}, M.~M., {L{\'e}pine}, S., \& {Bergeron}, P. 2013, \aj, 145, 136

\bibitem[{{Lindegren} {et~al.}(2021{\natexlab{a}}){Lindegren}, {Bastian},
  {Biermann}, {Bombrun}, {de Torres}, {Gerlach}, {Geyer}, {Hern{\'a}ndez},
  {Hilger}, {Hobbs}, {Klioner}, {Lammers}, {McMillan}, {Ramos-Lerate},
  {Steidelm{\"u}ller}, {Stephenson}, \& {van Leeuwen}}]{lindegren21b}
{Lindegren}, L., {Bastian}, U., {Biermann}, M., {et~al.} 2021{\natexlab{a}},
  \aap, 649, A4

\bibitem[{{Lindegren} {et~al.}(2021{\natexlab{b}}){Lindegren}, {Klioner},
  {Hern{\'a}ndez}, {Bombrun}, {Ramos-Lerate}, {Steidelm{\"u}ller}, {Bastian},
  {Biermann}, {de Torres}, {Gerlach}, {Geyer}, {Hilger}, {Hobbs}, {Lammers},
  {McMillan}, {Stephenson}, {Casta{\~n}eda}, {Davidson}, {Fabricius},
  {Gracia-Abril}, {Portell}, {Rowell}, {Teyssier}, {Torra}, {Bartolom{\'e}},
  {Clotet}, {Garralda}, {Gonz{\'a}lez-Vidal}, {Torra}, {Abbas}, {Altmann},
  {Anglada Varela}, {Balaguer-N{\'u}{\~n}ez}, {Balog}, {Barache}, {Becciani},
  {Bernet}, {Bertone}, {Bianchi}, {Bouquillon}, {Brown}, {Bucciarelli},
  {Busonero}, {Butkevich}, {Buzzi}, {Cancelliere}, {Carlucci}, {Charlot},
  {Cioni}, {Crosta}, {Crowley}, {del Peloso}, {del Pozo}, {Drimmel}, {Esquej},
  {Fienga}, {Fraile}, {Gai}, {Garcia-Reinaldos}, {Guerra}, {Hambly}, {Hauser},
  {Jan{\ss}en}, {Jordan}, {Kostrzewa-Rutkowska}, {Lattanzi}, {Liao}, {Licata},
  {Lister}, {L{\"o}ffler}, {Marchant}, {Masip}, {Mignard}, {Mints}, {Molina},
  {Mora}, {Morbidelli}, {Murphy}, {Pagani}, {Panuzzo}, {Pe{\~n}alosa Esteller},
  {Poggio}, {Re Fiorentin}, {Riva}, {Sagrist{\`a} Sell{\'e}s}, {Sanchez
  Gimenez}, {Sarasso}, {Sciacca}, {Siddiqui}, {Smart}, {Souami}, {Spagna},
  {Steele}, {Taris}, {Utrilla}, {van Reeven}, \& {Vecchiato}}]{lindegren21a}
{Lindegren}, L., {Klioner}, S.~A., {Hern{\'a}ndez}, J., {et~al.}
  2021{\natexlab{b}}, \aap, 649, A2

\bibitem[{{Lodders}(2003)}]{lodders03}
{Lodders}, K. 2003, \apj, 591, 1220

\bibitem[{{L{\'o}pez-Sanjuan} {et~al.}(2022{\natexlab{a}}){L{\'o}pez-Sanjuan},
  {Tremblay}, {Ederoclite}, {V{\'a}zquez Rami{\'o}}, {Carrasco}, {Varela},
  {Cenarro}, {Mar{\'\i}n-Franch}, {Civera}, {Daflon}, {G{\"a}nsicke}, {Gentile
  Fusillo}, {Jim{\'e}nez-Esteban}, {Alcaniz}, {Angulo},
  {Crist{\'o}bal-Hornillos}, {Dupke}, {Hern{\'a}ndez-Monteagudo}, {Moles}, \&
  {Sodr{\'e}}}]{clsj22pda}
{L{\'o}pez-Sanjuan}, C., {Tremblay}, P.~E., {Ederoclite}, A., {et~al.}
  2022{\natexlab{a}}, \aap, 658, A79

\bibitem[{{L{\'o}pez-Sanjuan} {et~al.}(2022{\natexlab{b}}){L{\'o}pez-Sanjuan},
  {Tremblay}, {Ederoclite}, {V{\'a}zquez Rami{\'o}}, {Cenarro},
  {Mar{\'\i}n-Franch}, {Varela}, {Akras}, {Guerrero}, {Jim{\'e}nez-Esteban},
  {Lopes de Oliveira}, {Chies-Santos}, {Fern{\'a}ndez-Ontiveros}, {Abramo},
  {Alcaniz}, {Ben{\'\i}tez}, {Bonoli}, {Carneiro}, {Crist{\'o}bal-Hornillos},
  {Dupke}, {Mendes de Oliveira}, {Moles}, {Sodr{\'e}}, \&
  {Taylor}}]{clsj22mjpaswd}
{L{\'o}pez-Sanjuan}, C., {Tremblay}, P.~E., {Ederoclite}, A., {et~al.}
  2022{\natexlab{b}}, \aap, 665, A151

\bibitem[{{L{\'o}pez-Sanjuan} {et~al.}(2021){L{\'o}pez-Sanjuan}, {Yuan},
  {V{\'a}zquez Rami{\'o}}, {Varela}, {Crist{\'o}bal-Hornillos}, {Tremblay},
  {Mar{\'\i}n-Franch}, {Cenarro}, {Ederoclite}, {Alfaro}, {Alvarez-Candal},
  {Daflon}, {Hern{\'a}n-Caballero}, {Hern{\'a}ndez-Monteagudo},
  {Jim{\'e}nez-Esteban}, {Placco}, {Tempel}, {Alcaniz}, {Angulo}, {Dupke},
  {Moles}, \& {Sodr{\'e}}}]{clsj21zsl}
{L{\'o}pez-Sanjuan}, C., {Yuan}, H., {V{\'a}zquez Rami{\'o}}, H., {et~al.}
  2021, \aap, 654, A61

\bibitem[{{Manser} {et~al.}(2024{\natexlab{a}}){Manser}, {G{\"a}nsicke},
  {Izquierdo}, {Swan}, {Najita}, {Rockosi}, {Carrillo}, {Kim}, {Xu}, {Dey},
  {Aguilar}, {Ahlen}, {Blum}, {Brooks}, {Claybaugh}, {Dawson}, {de la Macorra},
  {Doel}, {Gazta{\~n}aga}, {Gontcho A Gontcho}, {Honscheid}, {Kehoe}, {Kremin},
  {Landriau}, {Le Guillou}, {Levi}, {Li}, {Meisner}, {Miquel}, {Nie}, {Rezaie},
  {Rossi}, {Sanchez}, {Schubnell}, {Tarl{\'e}}, {Weaver}, {Zhou}, \&
  {Zou}}]{desi_edr_fca}
{Manser}, C.~J., {G{\"a}nsicke}, B.~T., {Izquierdo}, P., {et~al.}
  2024{\natexlab{a}}, \mnras, 531, L27

\bibitem[{{Manser} {et~al.}(2024{\natexlab{b}}){Manser}, {Izquierdo},
  {G{\"a}nsicke}, {Swan}, {Koester}, {Robert}, {Xu}, {Inight}, {Amroota},
  {Gentile Fusillo}, {Koposov}, {Kim}, {Dey}, {Allende Prieto}, {Aguilar},
  {Ahlen}, {Blum}, {Brooks}, {Claybaugh}, {Cooper}, {Dawson}, {de la Macorra},
  {Doel}, {Forero-Romero}, {Gazta{\~n}aga}, {Gontcho}, {Honscheid}, {Kisner},
  {Kremin}, {Lambert}, {Landriau}, {Le Guillou}, {Levi}, {Li}, {Meisner},
  {Miquel}, {Moustakas}, {Nie}, {Palanque-Delabrouille}, {Percival}, {Poppett},
  {Prada}, {Rezaie}, {Rossi}, {Sanchez}, {Schlafly}, {Schlegel}, {Schubnell},
  {Seo}, {Silber}, {Tarl{\'e}}, {Weaver}, {Zhou}, \& {Zou}}]{desi_edr_wd}
{Manser}, C.~J., {Izquierdo}, P., {G{\"a}nsicke}, B.~T., {et~al.}
  2024{\natexlab{b}}, arXiv e-prints, arXiv:2402.18641

\bibitem[{{Mar{\'{\i}}n-Franch} {et~al.}(2015){Mar{\'{\i}}n-Franch}, {Taylor},
  {Cenarro}, {Cristobal-Hornillos}, \& {Moles}}]{t80cam}
{Mar{\'{\i}}n-Franch}, A., {Taylor}, K., {Cenarro}, J., {Cristobal-Hornillos},
  D., \& {Moles}, M. 2015, in IAU General Assembly, Vol.~29, 2257381

\bibitem[{{McCleery} {et~al.}(2020){McCleery}, {Tremblay}, {Gentile Fusillo},
  {Hollands}, {G{\"a}nsicke}, {Izquierdo}, {Toonen}, {Cunningham}, \&
  {Rebassa-Mansergas}}]{40pcii}
{McCleery}, J., {Tremblay}, P.-E., {Gentile Fusillo}, N.~P., {et~al.} 2020,
  \mnras, 499, 1890

\bibitem[{{Mendes de Oliveira} {et~al.}(2019){Mendes de Oliveira}, {Ribeiro},
  {Schoenell}, {Kanaan}, {Overzier}, {Molino}, {Sampedro}, {Coelho}, {Barbosa},
  {Cortesi}, {Costa-Duarte}, {Herpich}, {Hernandez-Jimenez}, {Placco},
  {Xavier}, {Abramo}, {Saito}, {Chies-Santos}, {Ederoclite}, {Lopes de
  Oliveira}, {Gon{\c{c}}alves}, {Akras}, {Almeida}, {Almeida-Fernandes},
  {Beers}, {Bonatto}, {Bonoli}, {Cypriano}, {Vinicius-Lima}, {de Souza},
  {Fabiano de Souza}, {Ferrari}, {Gon{\c{c}}alves}, {Gonzalez},
  {Guti{\'e}rrez-Soto}, {Hartmann}, {Jaffe}, {Kerber}, {Lima-Dias}, {Lopes},
  {Menendez-Delmestre}, {Nakazono}, {Novais}, {Ortega-Minakata}, {Pereira},
  {Perottoni}, {Queiroz}, {Reis}, {Santos}, {Santos-Silva}, {Santucci},
  {Barbosa}, {Siffert}, {Sodr{\'e}}, {Torres-Flores}, {Westera}, {Whitten},
  {Alcaniz}, {Alonso-Garc{\'\i}a}, {Alencar}, {Alvarez-Candal}, {Amram},
  {Azanha}, {Barb{\'a}}, {Bernardinelli}, {Borges Fernandes}, {Branco},
  {Brito-Silva}, {Buzzo}, {Caffer}, {Campillay}, {Cano}, {Carvano}, {Castejon},
  {Cid Fernandes}, {Dantas}, {Daflon}, {Damke}, {de la Reza}, {de Melo de
  Azevedo}, {De Paula}, {Diem}, {Donnerstein}, {Dors}, {Dupke}, {Eikenberry},
  {Escudero}, {Faifer}, {Far{\'\i}as}, {Fernandes}, {Fernandes}, {Fontes},
  {Galarza}, {Hirata}, {Katena}, {Gregorio-Hetem},
  {Hern{\'a}ndez-Fern{\'a}ndez}, {Izzo}, {Jaque Arancibia}, {Jatenco-Pereira},
  {Jim{\'e}nez-Teja}, {Kann}, {Krabbe}, {Labayru}, {Lazzaro}, {Lima Neto},
  {Lopes}, {Magalh{\~a}es}, {Makler}, {de Menezes}, {Miralda-Escud{\'e}},
  {Monteiro-Oliveira}, {Montero-Dorta}, {Mu{\~n}oz-Elgueta}, {Nemmen}, {Nilo
  Castell{\'o}n}, {Oliveira}, {Ort{\'\i}z}, {Pattaro}, {Pereira}, {Quint},
  {Riguccini}, {Rocha Pinto}, {Rodrigues}, {Roig}, {Rossi}, {Saha}, {Santos},
  {Schnorr M{\"u}ller}, {Sesto}, {Silva}, {Smith Castelli}, {Teixeira},
  {Telles}, {Thom de Souza}, {Th{\"o}ne}, {Trevisan}, {de Ugarte Postigo},
  {Urrutia-Viscarra}, {Veiga}, {Vika}, {Vitorelli}, {Werle}, {Werner}, \&
  {Zaritsky}}]{splus}
{Mendes de Oliveira}, C., {Ribeiro}, T., {Schoenell}, W., {et~al.} 2019,
  \mnras, 489, 241

\bibitem[{{Montegriffo} {et~al.}(2023){Montegriffo}, {De Angeli}, {Andrae},
  {Riello}, {Pancino}, {Sanna}, {Bellazzini}, {Evans}, {Carrasco}, {Sordo},
  {Busso}, {Cacciari}, {Jordi}, {van Leeuwen}, {Vallenari}, {Altavilla},
  {Barstow}, {Brown}, {Burgess}, {Castellani}, {Cowell}, {Davidson}, {De
  Luise}, {Delchambre}, {Diener}, {Fabricius}, {Fr{\'e}mat}, {Fouesneau},
  {Gilmore}, {Giuffrida}, {Hambly}, {Harrison}, {Hidalgo}, {Hodgkin},
  {Holland}, {Marinoni}, {Osborne}, {Pagani}, {Palaversa}, {Piersimoni},
  {Pulone}, {Ragaini}, {Rainer}, {Richards}, {Rowell}, {Ruz-Mieres}, {Sarro},
  {Walton}, \& {Yoldas}}]{gaiadr3_calib}
{Montegriffo}, P., {De Angeli}, F., {Andrae}, R., {et~al.} 2023, \aap, 674, A3

\bibitem[{{O'Brien} {et~al.}(2023){O'Brien}, {Tremblay}, {Gentile Fusillo},
  {Hollands}, {G{\"a}nsicke}, {Koester}, {Pelisoli}, {Cukanovaite},
  {Cunningham}, {Doyle}, {Elms}, {Farihi}, {Hermes}, {Holberg}, {Jordan},
  {Klein}, {Kleinman}, {Manser}, {De Martino}, {Marsh}, {McCleery}, {Melis},
  {Nitta}, {Parsons}, {Raddi}, {Rebassa-Mansergas}, {Schreiber}, {Silvotti},
  {Steeghs}, {Toloza}, {Toonen}, {Torres}, {Weinberger}, \&
  {Zuckerman}}]{40pciii}
{O'Brien}, M.~W., {Tremblay}, P.~E., {Gentile Fusillo}, N.~P., {et~al.} 2023,
  \mnras, 518, 3055

\bibitem[{{O'Brien} {et~al.}(2024){O'Brien}, {Tremblay}, {Klein}, {Koester},
  {Melis}, {B{\'e}dard}, {Cukanovaite}, {Cunningham}, {Doyle}, {G{\"a}nsicke},
  {Gentile Fusillo}, {Hollands}, {McCleery}, {Pelisoli}, {Toonen},
  {Weinberger}, \& {Zuckerman}}]{40pciv}
{O'Brien}, M.~W., {Tremblay}, P.~E., {Klein}, B.~L., {et~al.} 2024, \mnras,
  527, 8687

\bibitem[{{Oke} \& {Gunn}(1983)}]{oke83}
{Oke}, J.~B. \& {Gunn}, J.~E. 1983, \apj, 266, 713

\bibitem[{{Pancino} {et~al.}(2012){Pancino}, {Altavilla}, {Marinoni},
  {Cocozza}, {Carrasco}, {Bellazzini}, {Bragaglia}, {Federici}, {Rossetti},
  {Cacciari}, {Balaguer N{\'u}{\~n}ez}, {Castro}, {Figueras}, {Fusi Pecci},
  {Galleti}, {Gebran}, {Jordi}, {Lardo}, {Masana}, {Mongui{\'o}},
  {Montegriffo}, {Ragaini}, {Schuster}, {Trager}, {Vilardell}, \&
  {Voss}}]{gaia_spss_i}
{Pancino}, E., {Altavilla}, G., {Marinoni}, S., {et~al.} 2012, \mnras, 426,
  1767

\bibitem[{{Pancino} {et~al.}(2021){Pancino}, {Sanna}, {Altavilla}, {Marinoni},
  {Rainer}, {Cocozza}, {Ragaini}, {Galleti}, {Bellazzini}, {Bragaglia},
  {Tessicini}, {Voss}, {Carrasco}, {Jordi}, {Harrison}, {De Angeli}, {Evans},
  \& {Fanari}}]{gaia_spss_v}
{Pancino}, E., {Sanna}, N., {Altavilla}, G., {et~al.} 2021, \mnras, 503, 3660

\bibitem[{Schwarz(1978)}]{schwarz78}
Schwarz, G. 1978, Ann. Statist., 6, 461

\bibitem[{{Starkenburg} {et~al.}(2017){Starkenburg}, {Martin}, {Youakim},
  {Aguado}, {Allende Prieto}, {Arentsen}, {Bernard}, {Bonifacio}, {Caffau},
  {Carlberg}, {C{\^o}t{\'e}}, {Fouesneau}, {Fran{\c{c}}ois}, {Franke},
  {Gonz{\'a}lez Hern{\'a}ndez}, {Gwyn}, {Hill}, {Ibata}, {Jablonka},
  {Longeard}, {McConnachie}, {Navarro}, {S{\'a}nchez-Janssen}, {Tolstoy}, \&
  {Venn}}]{pristine}
{Starkenburg}, E., {Martin}, N., {Youakim}, K., {et~al.} 2017, \mnras, 471,
  2587

\bibitem[{{Torres} {et~al.}(2023){Torres}, {Cruz}, {Murillo-Ojeda},
  {Jim{\'e}nez-Esteban}, {Rebassa-Mansergas}, {Solano}, {Camisassa}, {Raddi},
  \& {Doliguez Le Lourec}}]{torres23}
{Torres}, S., {Cruz}, P., {Murillo-Ojeda}, R., {et~al.} 2023, \aap, 677, A159

\bibitem[{{Tremblay} {et~al.}(2011){Tremblay}, {Bergeron}, \&
  {Gianninas}}]{tremblay11}
{Tremblay}, P.~E., {Bergeron}, P., \& {Gianninas}, A. 2011, \apj, 730, 128

\bibitem[{{Tremblay} {et~al.}(2020){Tremblay}, {Hollands}, {Gentile Fusillo},
  {McCleery}, {Izquierdo}, {G{\"a}nsicke}, {Cukanovaite}, {Koester}, {Brown},
  {Charpinet}, {Cunningham}, {Farihi}, {Giammichele}, {van Grootel}, {Hermes},
  {Hoskin}, {Jordan}, {Kepler}, {Kleinman}, {Manser}, {Marsh}, {de Martino},
  {Nitta}, {Parsons}, {Pelisoli}, {Raddi}, {Rebassa-Mansergas}, {Ren},
  {Schreiber}, {Silvotti}, {Toloza}, {Toonen}, \& {Torres}}]{40pci}
{Tremblay}, P.~E., {Hollands}, M.~A., {Gentile Fusillo}, N.~P., {et~al.} 2020,
  \mnras, 497, 130

\bibitem[{{Tremblay} {et~al.}(2013){Tremblay}, {Ludwig}, {Steffen}, \&
  {Freytag}}]{tremblay13}
{Tremblay}, P.~E., {Ludwig}, H.~G., {Steffen}, M., \& {Freytag}, B. 2013, \aap,
  559, A104

\bibitem[{{Turner} \& {Wyatt}(2020)}]{turner20}
{Turner}, S. G.~D. \& {Wyatt}, M.~C. 2020, \mnras, 491, 4672

\bibitem[{{Vincent} {et~al.}(2023){Vincent}, {Barstow}, {Jordan}, {Mander},
  {Bergeron}, \& {Dufour}}]{vincent24}
{Vincent}, O., {Barstow}, M.~A., {Jordan}, S., {et~al.} 2023, arXiv e-prints,
  arXiv:2308.05572

\bibitem[{{Wenger} {et~al.}(2000){Wenger}, {Ochsenbein}, {Egret}, {Dubois},
  {Bonnarel}, {Borde}, {Genova}, {Jasniewicz}, {Lalo{\"e}}, {Lesteven}, \&
  {Monier}}]{simbad}
{Wenger}, M., {Ochsenbein}, F., {Egret}, D., {et~al.} 2000, \aaps, 143, 9

\bibitem[{{Zuckerman} {et~al.}(2003){Zuckerman}, {Koester}, {Reid}, \&
  {H{\"u}nsch}}]{zuckerman03}
{Zuckerman}, B., {Koester}, D., {Reid}, I.~N., \& {H{\"u}nsch}, M. 2003, \apj,
  596, 477

\bibitem[{{Zuckerman} {et~al.}(2010){Zuckerman}, {Melis}, {Klein}, {Koester},
  \& {Jura}}]{zuckerman10}
{Zuckerman}, B., {Melis}, C., {Klein}, B., {Koester}, D., \& {Jura}, M. 2010,
  \apj, 722, 725

\end{thebibliography}

\begin{appendix}
\section{Catalog of white dwarf parameters}\label{app:data}
\begin{table*} 
\caption{White dwarf catalog of atmospheric parameters and composition.}
\label{tab:catalog}
\centering 
        \begin{tabular}{c c l}
        \hline\hline\rule{0pt}{3ex} 
        Heading     & Units  &  Description   \\
        \hline\rule{0pt}{3ex}
      \!TILE-ID     &  $\cdots$       & Identifier of the J-PLUS Tile image in the $r$ band where the object was detected. \\ 
        NUMBER       &  $\cdots$       & Number identifier assigned by \texttt{SExtractor} for the object in the $r$-band image. \\
        RAdeg        &  deg            & Right ascension (J2000). \\
        DEdeg        &  deg            & Declination (J2000). \\
        pca  &  $\cdots$       & Probability of having \ion{Ca}{ii} H+K absorption.\\
        EWJ0395 & $\AA$      & Equivalent width of the $J0395$ passband.\\
        e\_EWJ0395 & $\AA$   & Error in the equivalent width of the $J0395$ passband.\\
        Teff         &  K              & Effective temperature ($T_{\rm eff}$) from J-PLUS photometry but $J0395$ passband.\\ 
        e\_Teff      &  K              & Uncertainty in Teff.\\
        Mass         &  $M_{\odot}$    & Mass from J-PLUS photometry but $J0395$ passband.\\ 
        e\_Mass   &  $M_{\odot}$       & Uncertainty in mass.\\ 
        \hline 
\end{tabular}
\end{table*}
In this Appendix, we describe the data of the $\nwd$ white dwarfs analyzed in the present paper. The gathered information is in the J-PLUS database\footnote{\url{http://archive.cefca.es/catalogues/jplus-dr2/tap_async.html}} at the table \texttt{jplus.WhiteDwarf}, also accessible through virtual observatory table access protocol (TAP) service. The catalog is also provided at CDS. The description of the columns is provided in Table.~\ref{tab:catalog}.

The information in the catalog is also accessible with the J-PLUS explorer for individual objects, including a graphical view of the best-fit solution as in Fig.~\ref{fig:example}.
\end{appendix}

\end{document}